\documentclass[prd,twocolumn,twoside,preprintnumbers,superscriptaddress,nofootinbib]{revtex4}
\usepackage{amsmath,slashed}
\usepackage{graphicx,graphics}
\usepackage{dcolumn}
\usepackage[hyperfootnotes=false]{hyperref}
\usepackage{xspace}

\newcommand{\Fpi}{F_\pi}
\newcommand{\mpi}{M_{\pi}}
\newcommand{\mpii}{M_{\pi^0}}
\newcommand{\Order}{\mathcal{O}}
\newcommand{\MeV}{\,\text{MeV}}
\newcommand{\GeV}{\,\text{GeV}}
\newcommand{\TeV}{\,\text{TeV}}
\newcommand{\BR}{\text{BR}}
\newcommand{\beq}{\begin{equation}}
\newcommand{\eeq}{\end{equation}}
\newcommand{\diff}{\text{d}}
\newcommand{\sis}{s_\text{is}}
\newcommand{\siv}{s_\text{iv}}
\newcommand{\sthr}{s_\text{thr}}
\newcommand{\sm}{s_\text{m}}
\newcommand{\grhop}{g_\text{eff}}
\newcommand{\Mrhop}{M_\text{eff}}

\renewcommand{\Re}{\text{Re}\,}
\renewcommand{\Im}{\text{Im}\,}
\newcommand{\A}{\mathcal{A}}
\newcommand{\Lagr}{\mathcal{L}}

\allowdisplaybreaks[1]

\begin{document}


\title{Improved Standard-Model prediction for $\boldsymbol{\pi^0\to e^+e^-}$}

\author{Martin Hoferichter}
\affiliation{Albert Einstein Center for Fundamental Physics, Institute for Theoretical Physics, University of Bern, Sidlerstrasse 5, 3012 Bern, Switzerland}
\author{Bai-Long Hoid}
\affiliation{Albert Einstein Center for Fundamental Physics, Institute for Theoretical Physics, University of Bern, Sidlerstrasse 5, 3012 Bern, Switzerland}
\author{Bastian Kubis}
\affiliation{Helmholtz-Institut f\"ur Strahlen- und Kernphysik and Bethe Center for Theoretical Physics, Universit\"at Bonn, 53115 Bonn, Germany}
\author{Jan L\"udtke}
\affiliation{Helmholtz-Institut f\"ur Strahlen- und Kernphysik and Bethe Center for Theoretical Physics, Universit\"at Bonn, 53115 Bonn, Germany}
\affiliation{Faculty of Physics, University of Vienna, Boltzmanngasse 5, 1090 Vienna, Austria}

\begin{abstract}
We present an improved Standard-Model (SM) prediction for the dilepton decay of the neutral pion. 
The loop amplitude is determined by 
the pion transition form factor for $\pi^0\to\gamma^*\gamma^*$, for which we employ a dispersive representation that incorporates both space-like and time-like data as well as short-distance constraints. The resulting SM branching fraction, $ \BR[\pi^0\to e^+e^-]=6.25(3)\times 10^{-8}$ , sharpens constraints on physics beyond the SM, including pseudoscalar and axial-vector mediators. 
\end{abstract} 

\maketitle

\section{Introduction}

The decay of the neutral pion proceeds almost exclusively into two photons, with the decay mediated by the Wess--Zumino--Witten anomaly~\cite{Wess:1971yu,Witten:1983tw}. The decay width 
\beq
\Gamma[\pi^0\to\gamma\gamma]=\frac{\pi\alpha^2\mpii^3}{4}F_{\pi\gamma\gamma}^2
\eeq
depends on the pion transition form factor (TFF) at zero momentum transfer, $F_{\pi\gamma\gamma}=F_{\pi^0\gamma^*\gamma^*}(0,0)$, which in turn is determined by a low-energy theorem~\cite{Adler:1969gk,Bell:1969ts,Bardeen:1969md}
\beq
F_{\pi\gamma\gamma}=\frac{1}{4\pi^2 F_\pi}=0.2745(3)\GeV^{-1},  
\eeq
in terms of the pion decay constant $F_\pi=92.28(10)\MeV$~\cite{Zyla:2020zbs}. This prediction agrees extremely well with experiment, $F_{\pi\gamma\gamma}=0.2754(21)\GeV^{-1}$~\cite{Larin:2020bhc}, despite the fact that at this level higher-order corrections are expected~\cite{Bijnens:1989jb,Ananthanarayan:2002kj,Goity:2002nn,Kampf:2009tk}.
The second-most important decay channel is the Dalitz decay $\pi^0\to e^+e^-\gamma$. Combining the radiative corrections from Ref.~\cite{Husek:2015sma} with phenomenological input on the slope of the TFF~\cite{Hoferichter:2014vra,
Hoferichter:2018dmo,Hoferichter:2018kwz,Masjuan:2012wy,Masjuan:2017tvw,Behrend:1990sr,Adlarson:2016ykr,TheNA62:2016fhr} gives~\cite{Husek:2018qdx}  
\beq
\BR[\pi^0\to\gamma\gamma]=98.8131(6)\%,
\eeq
in agreement with but more precise than the direct measurement $\BR[\pi^0\to\gamma\gamma]=98.823(34)\%$~\cite{Zyla:2020zbs,Samios:1961zz,Schardt:1980qd,Beddall:2008zza}. The decay $\pi^0\to 2(e^+e^-)$ is suppressed by another factor of $\alpha$ with respect to the Dalitz decay, leading to $\BR[\pi^0\to 2(e^+e^-)]=3.26(18)\times 10^{-5}$~\cite{Abouzaid:2008cd}. 

\begin{figure}[t]
	\includegraphics[width=\linewidth]{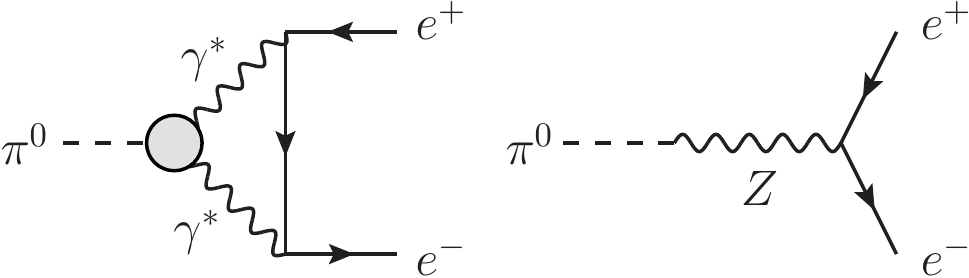}
	\caption{SM contributions to $\pi^0\to e^+e^-$, with the dominant $\pi^0\to\gamma^*\gamma^*$ diagram (left) and a small correction from $Z$ exchange (right). The gray blob refers to the pion TFF. } 
	\label{fig:diagrams}
\end{figure}

Here, we are interested in the rare decay $\pi^0\to e^+e^-$, whose dominant contribution in the SM arises from the loop diagram shown in Fig.~\ref{fig:diagrams}. Apart from a loop factor, there is yet another chiral suppression, 
which is partly lifted by logarithmic corrections. Altogether, this leads to the 
scaling~\cite{Drell:1959}
\beq
\label{scaling}
\BR[\pi^0\to e^+e^-]\sim 
\Big(\frac{\alpha}{\pi}\Big)^2\frac{m_e^2}{\mpii^2} \pi^2 \log^2\frac{m_e}{\mpii}\sim \Order\big(10^{-8}\big),
\eeq
and due to the corresponding suppression the decay has been suggested early on 
to search for physics beyond the SM (BSM)~\cite{Soni:1974aw}. Since the 
interference with the $Z$-boson contribution, see second diagram in 
Fig.~\ref{fig:diagrams}, is suppressed by another two orders of 
magnitude~\cite{Arnellos:1981bk,Masjuan:2015lca,Masjuan:2015cjl}, sensitivity to 
BSM degrees of freedom in general requires a precision measurement of 
$\BR[\pi^0\to e^+e^-]$, unless the BSM contribution is enhanced in one way or 
another. Such enhancement could originate from avoiding the chiral suppression 
in Eq.~\eqref{scaling} via pseudoscalar operators, or by considering light 
degrees of freedom, such as axial-vector $Z'$ 
bosons~\cite{Kahn:2007ru,Kahn:2016vjr} or 
axion-like 
particles~\cite{Chang:2008np,Andreas:2010ms,Bauer:2017ris,Alves:2017avw,
Altmannshofer:2019yji,Bauer:2021mvw}.   

The current best measurement by the KTeV experiment constrains $\BR[\pi^0\to e^+e^-]$ at the level of $5\%$~\cite{Abouzaid:2006kk}, but the interpretation is complicated by the fact that the result is provided with a cut on the dilepton invariant mass, which needs to be extrapolated to obtain the full branching fraction. Using the latest radiative corrections from Refs.~\cite{Vasko:2011pi,Husek:2014tna}, one finds
\beq
\label{KTeV_BR}
\BR[\pi^0\to e^+e^-]\big|_\text{KTeV}=6.85(27)(23)\times 10^{-8},
\eeq
significantly lower than the extrapolation $
\BR[\pi^0\to e^+e^-]=7.48(38)\times 10^{-8}$ given in Ref.~\cite{Abouzaid:2006kk} based on the radiative corrections from Ref.~\cite{Bergstrom:1982wk} (see also Ref.~\cite{Dorokhov:2008qn}), with the difference due to the assumption of a point-like $\pi^0\to e^+e^-$ vertex in Ref.~\cite{Bergstrom:1982wk}.\footnote{
We used the total correction $\delta=-6.0(2)\%$ in Eq.~\eqref{KTeV_BR}, in line 
with the low-energy constant $\chi^{(\text{r})}(\mu=0.77\GeV)=2.69(10)$ that 
corresponds to our result for the pion 
TFF~\cite{Vasko:2011pi,Husek:2014tna,Husek:2021}, see below. Note that precisely due to the potentially complicated
dependence on kinematical cuts, we follow the convention to subtract the
radiative corrections from the experimental result and use the leading
order in QED as the reference point for comparison between theory and
experiment.}
 
In the SM prediction, the imaginary part of the amplitude due to the 
$\gamma\gamma$ cut is determined model-independently in terms of 
$\Gamma[\pi^0\to\gamma\gamma]$, leading to a unitarity bound of $\BR[\pi^0\to 
e^+e^-]>4.69\times 10^{-8}$~\cite{Berman:1960zz,Pratap:1972tb}. To obtain the 
real part,
additional information needs to be provided on the TFF, see Ref.~\cite{Gan:2020aco} for a review. Chiral perturbation theory (ChPT) only allows one to relate the dilepton decays of $\pi^0$ and $\eta^{(\prime)}$, but cannot predict $\BR[\pi^0\to e^+e^-]$ itself~\cite{Savage:1992ac,GomezDumm:1998gw}.  
Further approaches that have been pursued instead include vector-meson-dominance 
TFFs~\cite{Ametller:1993we,Knecht:1999gb,Silagadze:2006rt,Husek:2015wta} and a 
dispersion relation in the pion mass 
squared~\cite{Bergstrom:1983ay,Ametller:1983ec,Dorokhov:2007bd,Dorokhov:2008cd,
Dorokhov:2009xs} . However, we stress that such a dispersion relation is model 
dependent, as the TFF for unphysical masses is not observable, and the input for 
the imaginary part is typically restricted to the $\gamma\gamma$ cut. 

More recent SM predictions include $\BR[\pi^0\to e^+e^-]=6.23(5)\times 10^{-8}$~\cite{Masjuan:2015lca} based on Canterbury approximants and 
$\BR[\pi^0\to e^+e^-]=6.22(3)\times 10^{-8}$~\cite{Weil:2017knt} (excluding $Z$ exchange)
using Dyson--Schwinger equations~\cite{Eichmann:2017wil}. The Canterbury expansion relies on space-like data for the pion TFF~\cite{Behrend:1990sr,Gronberg:1997fj,Aubert:2009mc,Uehara:2012ag} and is, in principle, systematically improvable, but in practice restricted due to the available data, especially lack thereof in the doubly-virtual direction, while for the Dyson--Schwinger approach a complete estimate of the truncation uncertainties is challenging. 
 The $\pi^0\to e^+e^-$ decay is also becoming amenable to calculations in lattice QCD~\cite{Christ:2020dae}.  

In this Letter, we present a SM prediction that is based on a dispersive representation of the pion TFF first developed in the context of the pion-pole contribution~\cite{Hoferichter:2014vra,
Hoferichter:2018dmo,Hoferichter:2018kwz} in a dispersive approach to hadronic light-by-light (HLbL) scattering~\cite{Hoferichter:2013ama,Colangelo:2014dfa,Colangelo:2014pva,Colangelo:2015ama,Colangelo:2017qdm,Colangelo:2017fiz,Danilkin:2021icn}, with further applications to hadronic vacuum polarization~\cite{Hoferichter:2019mqg,Hoid:2020xjs}. In the dispersive approach presented here we are able to implement constraints from all available low-energy data, including the time-like region, to predict the doubly-virtual behavior from singly-virtual data, and to ensure a smooth matching to short-distance constraints. The resulting SM prediction, which is as precise as we believe can currently be achieved with data-driven methods, is then used to sharpen some of the constraints that can be extracted from the comparison to the KTeV measurement. We also clarify some technical points in the calculation of the SM amplitude, and show that a Wick rotation to space-like momenta is possible once a double-spectral representation is employed for the pion TFF.

\section{Pion transition form factor}

The pion TFF is defined by the matrix element
of two electromagnetic currents $j_\mu(x)$
\begin{align}
  \label{eq:defpiTFF}
  & i\int \diff^4x \, e^{iq_1\cdot x} \, \langle 0 \vert T \{ j_\mu(x) \, j_\nu(0)\} \vert \pi^0(q_1+q_2) \rangle
  \notag \\
  & = \epsilon_{\mu\nu\alpha\beta} \, q_1^\alpha \, q_2^\beta \, F_{\pi^0\gamma^*\gamma^*}(q_1^2,q_2^2),
\end{align}
where we follow the sign conventions of Refs.~\cite{Colangelo:2019lpu,Colangelo:2019uex,Hoferichter:2020lap} to ensure consistency with the short-distance constraints and the $Z$-boson contribution. This form factor has been studied in great detail in the context of HLbL scattering~\cite{Masjuan:2017tvw,Hoferichter:2018dmo,Hoferichter:2018kwz,Gerardin:2019vio}, the key difference being that in this case the loop integral can be Wick-rotated to space-like momenta for an arbitrary TFF~\cite{Knecht:2001qf}. In the case of $\pi^0\to e^+e^-$ the analogous master formula becomes more intricate, so before turning to this application we first describe the representation we will use for the normalized TFF $\tilde F_{\pi^0\gamma^*\gamma^*}(q_1^2,q_2^2)=F_{\pi^0\gamma^*\gamma^*}(q_1^2,q_2^2)/F_{\pi\gamma\gamma}$ in the following. 
We use the decomposition 
\beq
\label{TFF_final}
\tilde F_{\pi^0\gamma^*\gamma^*}=\tilde F_{\pi^0\gamma^*\gamma^*}^\text{disp}+\tilde F_{\pi^0\gamma^*\gamma^*}^\text{eff}+\tilde F_{\pi^0\gamma^*\gamma^*}^\text{asym},
\eeq
where the dispersive term accounts for the low-energy singularities, extracted from data on $e^+e^-\to 2\pi,3\pi$; the second term parameterizes the small effect from higher intermediate states and high-energy contributions, it enforces the correct normalization and is further constrained by high-energy space-like data; and the third term implements the remaining short-distance constraints as expected from perturbative QCD. 

In practice, the dispersive part is written as a double-spectral representation (exploiting the absence of anomalous thresholds in this case~\cite{Lucha:2006vc,Colangelo:2015ama})
\begin{align}
\label{low_energy}
\tilde F^\text{disp}_{\pi^0\gamma^*\gamma^*}(q_1^2,q_2^2)&=
\frac{1}{\pi^2} \int_{4M_\pi^2}^{\siv} \diff x \int_{\sthr}^{\sis} \diff y  \frac{\tilde\rho(x,y)}{\big(x-q_1^2\big)\big(y-q_2^2\big)}\notag\\
&+(q_1\leftrightarrow q_2),\\
\tilde \rho(x,y)&=\frac{q_\pi^3(x)}{12\pi\sqrt{x} F_{\pi\gamma\gamma}}\Im \Big[\big(F_\pi^{V}(x)\big)^*f_1(x,y)\Big],\notag 
\end{align}  
with $q_\pi(s)=\sqrt{s/4-\mpi^2}$, and $\sthr=9\mpi^2$ or $\mpii^2$ depending on 
whether isospin-breaking corrections are included. The double-spectral density 
$\tilde \rho(x,y)$ is determined by the electromagnetic form factor of the pion, 
$F_\pi^V$, and the $P$-wave amplitude for $\gamma^*\to 3\pi$, $f_1$. The former 
is known very precisely from $e^+e^-\to 2\pi$ data (see, e.g., 
Refs.~\cite{Colangelo:2018mtw,Colangelo:2020lcg}), while the latter can be 
obtained from a solution of Khuri--Treiman equations~\cite{Khuri:1960zz}, with 
free parameters determined from $e^+e^-\to 3\pi$ 
data~\cite{Hoferichter:2014vra, Hoferichter:2018dmo,Hoferichter:2018kwz}. The 
integration cutoffs  are varied between $1.8$ and $2.5\GeV$, which, together 
with the variations of the $\pi\pi$ phase shifts and the conformal polynomial in 
the partial wave $f_1$, defines the dispersive contribution to the uncertainty 
estimate. 

The unsubtracted dispersion relation~\eqref{low_energy} only saturates the normalization at the level of $90\%$, with the remainder restored by an effective-pole contribution, 
\beq
\tilde F_{\pi^0\gamma^*\gamma^*}^\text{eff}(q_1^2,q_2^2)=\grhop\frac{\Mrhop^4}{(\Mrhop^2-q_1^2)(\Mrhop^2-q_2^2)},
\eeq
that accounts for higher intermediate states beyond $2\pi$, $3\pi$ as well as the high-energy part of the integrals. The coupling $\grhop$ follows from the normalization, while the mass scale $\Mrhop$ is determined from a fit to the singly-virtual space-like data~\cite{Behrend:1990sr,Gronberg:1997fj,Aubert:2009mc,Uehara:2012ag} with $Q^2 > 5\GeV^2$, to ensure that the low-energy properties remain unaffected. The resulting value of $\Mrhop$ lies in the range $1.5$--$2\GeV$, with an uncertainty dominated by the systematic tension between the BaBar data~\cite{Aubert:2009mc} and the other data sets, as well as the Brodsky--Lepage (BL) limit~\cite{Lepage:1979zb,Lepage:1980fj}.  

\begin{figure}[t]
	\includegraphics[width=\linewidth]{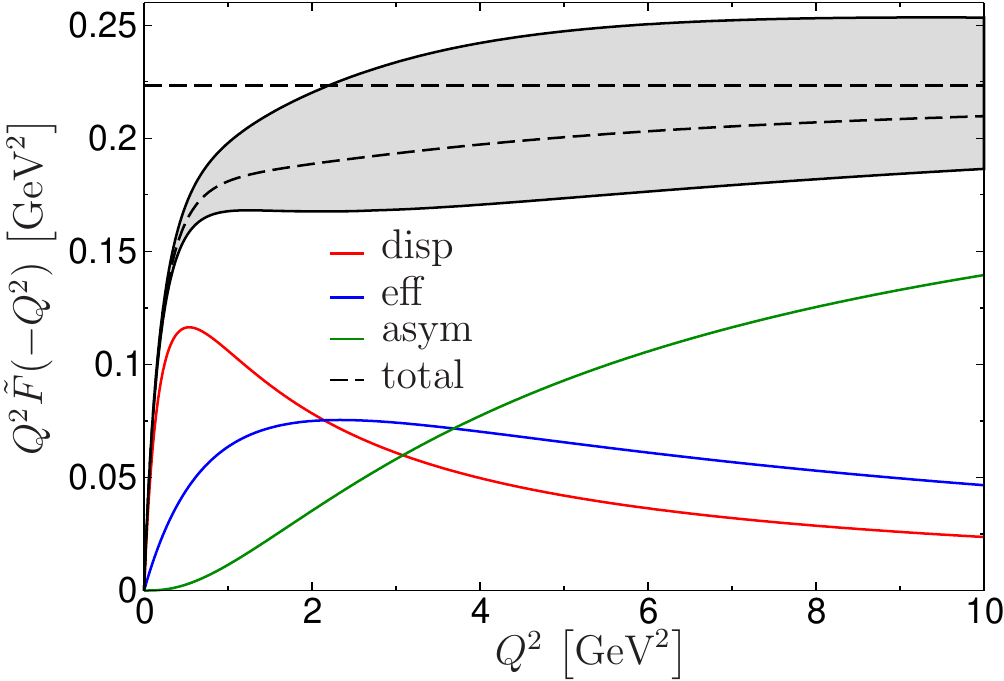}
	\caption{Dispersive, effective-pole, and asymptotic contributions to the pion TFF, using the representation from Eq.~\eqref{Fspace} with $\tilde k^2\to-Q^2$~\cite{Supparxiv}. This form factor $\tilde F(-Q^2)$ defines a single-variable function that is closely related to the input required for a space-like evaluation of the loop integral, for which $\tilde F_{\pi^0\gamma^*\gamma^*}(-Q^2,-Q^2)$ is not sufficient. In particular, due to the $\mpii^2$ corrections, the form factor $\tilde F(-Q^2)$ is not normalized exactly to unity at $Q^2=0$. The gray band indicates our uncertainty estimate, the horizontal dashed line the asymptotic value $2F_\pi/(3F_{\pi\gamma\gamma})$.} 
	\label{fig:TFF}
\end{figure}

Finally, the asymptotic contribution
\beq
\label{asym}
\tilde F_{\pi^0\gamma^*\gamma^*}^\text{asym}(q_1^2,q_2^2)
 = \frac{2\Fpi}{F_{\pi\gamma\gamma}}\int_{\sm}^\infty \diff x \frac{q_1^2q_2^2}{(x-q_1^2)^2(x-q_2^2)^2}
\eeq
ensures the correct asymptotic behavior for nonvanishing virtualities, and has been derived by expressing the short-distance constraints in terms of a dispersion relation~\cite{Khodjamirian:1997tk,Hoferichter:2018dmo,Hoferichter:2018kwz,Zanke:2021wiq}. The matching point is chosen as $\sm=1.7(3)\GeV^2$, in accordance with expectations from light-cone sum rules~\cite{Khodjamirian:1997tk,Agaev:2010aq,Mikhailov:2016klg}. The resulting TFF that emerges from the sum in Eq.~\eqref{TFF_final} is illustrated in Fig.~\ref{fig:TFF} for the kinematic configuration most relevant for $\pi^0\to e^+e^-$, demonstrating that our representation smoothly connects the various constraints on the pion TFF.

\section{SM prediction for $\boldsymbol{\pi^0\to e^+e^-}$}

The normalized branching fraction for $\pi^0\to e^+e^-$,
\beq
\frac{\BR[\pi^0\to e^+e^-]}{\BR[\pi^0\to\gamma\gamma]}=2\sigma_e(q^2)\Big(\frac{\alpha}{\pi}\Big)^2\frac{m_e^2}{\mpii^2}\big|\A(q^2)\big|^2,
\eeq
is typically expressed in terms of
the reduced amplitude 
\begin{align}
\label{Aloop}
\A(q^2)&=\frac{2i}{\pi^2q^2}\int\diff^4 k\frac{q^2k^2-(q\cdot k)^2}{k^2(q-k)^2[(p-k)^2-m_e^2]}\notag\\
&\times\tilde F_{\pi^0\gamma^*\gamma^*}\big(k^2,(q-k)^2\big),
\end{align}
where $q^2=\mpii^2$ and $p$ is the momentum of the outgoing electron. The only imaginary part arises from the $\gamma\gamma$ cut, which leads to 
\begin{align}
\Im\A(q^2)&=\frac{\pi}{2\sigma_e(q^2)}\log \big[y_e(q^2)\big]=-17.52,\notag\\
y_e(q^2)&=\frac{1-\sigma_e(q^2)}{1+\sigma_e(q^2)},\quad \sigma_e(q^2)=\sqrt{1-\frac{4m_e^2}{q^2}},
\end{align}
and defines the unitarity bound given above.

To obtain $\Re\A(q^2)$ we need to perform the integral~\eqref{Aloop} for our representation of the pion TFF. For the dispersive part one may write 
\begin{align}
 \A^\text{disp}(q^2)&=\frac{2}{\pi^2}
 \int_{4M_\pi^2}^{\siv} \diff x \int_{\sthr}^{\sis} \diff y\frac{\tilde \rho(x,y)}{x y} K(x,y),
\end{align}
where the integration kernel
\begin{align}
 K(x,y)&=\frac{2i}{\pi^2 q^2}\int \diff^4 k
 \frac{q^2k^2-(q\cdot k)^2}{k^2(q-k)^2[(p-k)^2-m_e^2]}\notag\\
 &\times \frac{x y}{(k^2-x)[(q-k)^2-y]}
\end{align}
can be evaluated based on standard loop functions, see Refs.~\cite{Masjuan:2015cjl,Luedtke:2016,Hoid:2020qij} and App.~\ref{app:loop_functions}. The effective-pole contribution follows from $x=y=\Mrhop^2$, and a similar decomposition can be derived for $\A^\text{asym}(q^2)$. The numerical integration over the double-spectral function requires a stable implementation of $K(x,y)$ over a wide parameter range, especially in view of the singularity structure of $f_1(x,y)$ that needs to be properly taken into account. As discussed in App.~\ref{app:loop_functions}, we verified the numerical stability by comparing several different methods, in particular, a Wick rotation to space-like momenta. Such a Wick rotation is not possible for a completely general TFF, but does apply for the double-spectral representation. Combining first the photon propagators, the angular part of the integral can be performed analytically with the method of Gegenbauer polynomials~\cite{Rosner:1967zz,Levine:1974xh,Levine:1979uz}, leaving an integration over a space-like modulus. In practice, however, we do not use this implementation of the loop functions, as it proves numerically less viable than other methods, including semi-analytic expressions in terms of polylogarithms~\cite{tHooft:1978jhc} and the implementation from \textsc{LoopTools}~\cite{Hahn:1998yk}.     

In the end, we find for the long-range contribution
\beq
\label{disp_result}
\Re\A(q^2)\big|_{\gamma^*\gamma^*}=
10.16(5)_\text{disp}(8)_\text{BL} (2)_\text{asym},
\eeq
with an uncertainty dominated by the systematic tensions around the BL limit.\footnote{For comparison we quote $\A(q^2)\big|_{\gamma^*\gamma^*}=10.10(3)-17.45(1)i$~\cite{Weil:2017knt} and $\Re\A(q^2)\big|_{\gamma^*\gamma^*}=10.08(16)$ reconstructed from the decay rate given in Ref.~\cite{Masjuan:2015lca}.} The full number decomposes as $10.16=9.18_\text{disp}+1.08_\text{eff}-0.10_\text{asym}$ according to the three terms in Eq.~\eqref{TFF_final}, reflecting the hierarchy expected from Fig.~\ref{fig:TFF}. Matching to ChPT
\begin{align}
 \Re\A(q^2)|_\text{ChPT}&=\frac{\text{Li}_2[-y_e(q^2)]+\frac{1}{4}\log^2\big[y_e(q^2)]+\frac{\pi^2}{12}}{\sigma_e(q^2)}\notag\\
 &+3\log\frac{m_e}{\mu}-\frac{5}{2}+\chi^{(\text{r})}(\mu)
\end{align}
then also determines the low-energy constant $\chi^{(\text{r})}(\mu=0.77\GeV)=2.69(10)$ (see, e.g., Refs.~\cite{Hoferichter:2018kwz,Husek:2014tna} for the conventions).

At this level of precision the contribution from the asymptotic region thus needs to be included, as does the $Z$-boson exchange~\cite{Masjuan:2015cjl}
\beq
\label{Zexchange}
\Re\A(q^2)\big|_Z=-\frac{F_\pi G_F}{\sqrt{2}\,\alpha^2F_{\pi\gamma\gamma}}=-0.05(0). 
\eeq
Adding both contributions, we
obtain the SM prediction
\begin{align}
\label{SM}
 \Re\A(q^2)\big|_\text{SM}&=10.11(10),\notag\\
 \BR[\pi^0\to e^+e^-]\big|_\text{SM}&=6.25(3)\times 10^{-8},
\end{align}
in a mild $1.8\sigma$ tension with the KTeV measurement~\eqref{KTeV_BR}. 
In particular, the latter implies
\begin{align}
\label{KTeV}
 \Re\A(q^2)\big|_\text{KTeV}&=11.89^{+0.94}_{-1.02},
\end{align}
which, in comparison to Eq.~\eqref{SM}, can directly be used to constrain effects beyond the SM.

\begin{figure*}[t]
	\includegraphics[width=0.48\linewidth]{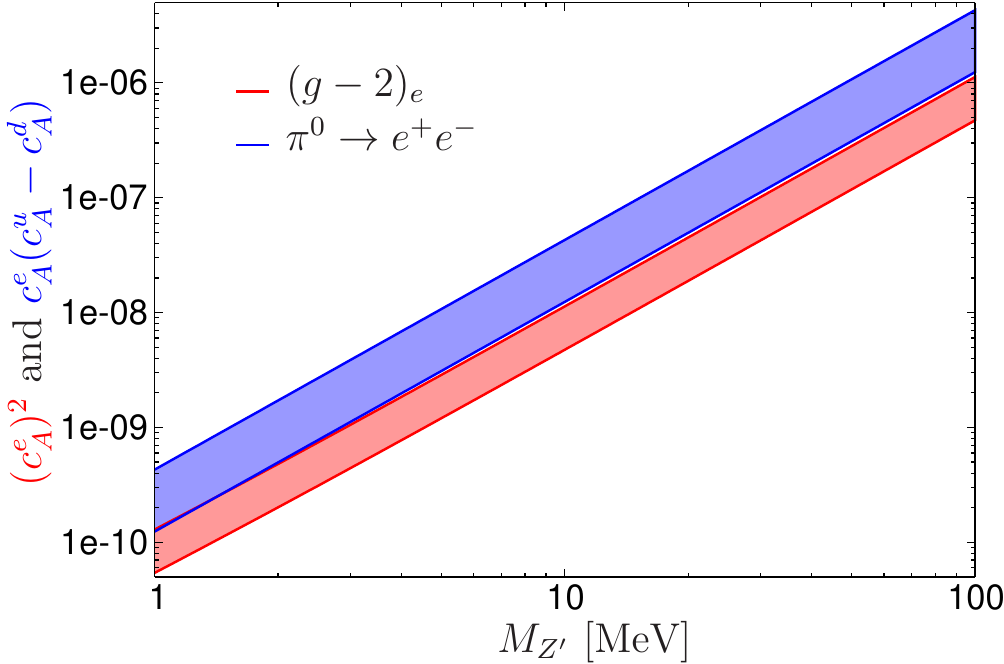}
	\includegraphics[width=0.48\linewidth]{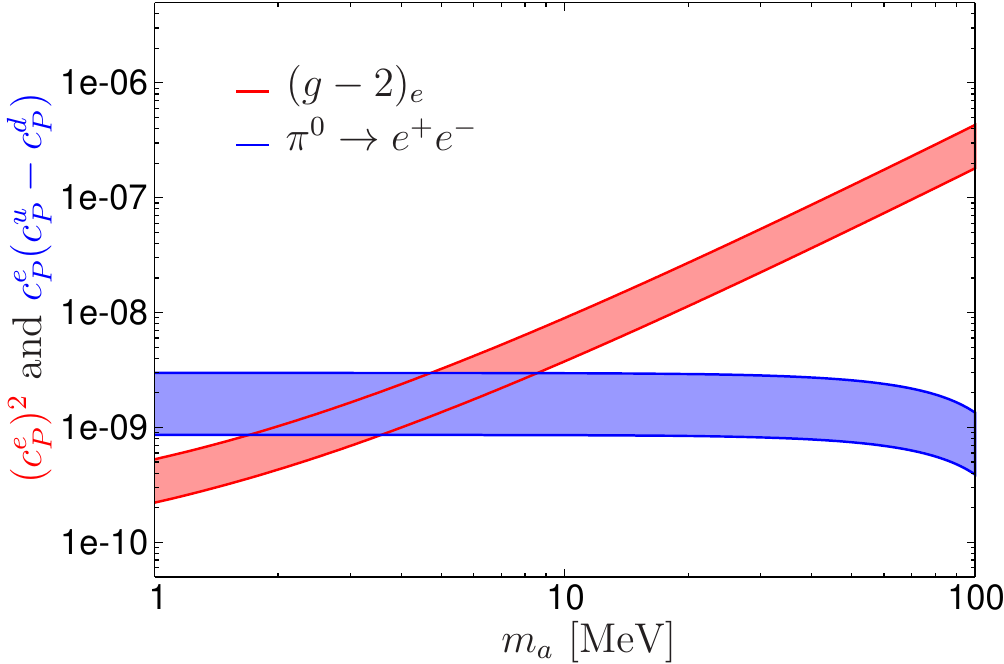}
	\caption{$1\sigma$ parameter ranges preferred by $\Delta a_e[\text{Cs}]$ (red) and $\pi^0\to e^+e^-$ (blue) on the couplings of light axial-vector (left) and pseudoscalar (right) mediators.} 
	\label{fig:bands}
\end{figure*}

\section{Constraints on BSM physics}

The comparison between our improved SM prediction~\eqref{SM} and the KTeV measurement~\eqref{KTeV} sharpens the constraints on physics beyond the SM. Writing new short-range interactions of
axial-vector and pseudoscalar type as
\beq
\Lagr_\text{BSM}^{(1)}=C_A \bar q\frac{\tau^3}{2}\gamma^\mu\gamma_5 q\, \bar e\gamma_\mu\gamma_5 e
+C_P \bar q\frac{\tau^3}{2}i\gamma_5 q\, \bar e i\gamma_5 e,
\eeq
with $q=(u,d)^T$, we obtain
\beq
\Re \A(q^2)|_\text{BSM}=-\frac{F_\pi}{\alpha^2 F_{\pi\gamma\gamma}}\bigg(C_A+\frac{\mpii^2}{4m_e\hat m}C_P\bigg),
\eeq
where $\hat m=(m_u+m_d)/2$, cf.\ also Refs.~\cite{Masjuan:2015lca,Masjuan:2015cjl,Sanchez-Puertas:2018tnp}. In particular, integrating out $Z$ exchange in the SM gives $C_A=G_F/\sqrt{2}$, in agreement with Eq.~\eqref{Zexchange}. 

The limits derived from $\pi^0\to e^+e^-$ are
\beq
\label{CACP}
C_A=(-280)^{+160}_{-150}\TeV^{-2},\quad 
C_P=(-0.108)^{+0.062}_{-0.057}\TeV^{-2},
\eeq
where the pseudoscalar coefficient has been evaluated at the $\overline{\text{MS}}$ scale $\mu=2\GeV$ using
$\hat m=3.4\MeV$~\cite{Aoki:2019cca,McNeile:2010ji,Durr:2010vn,Carrasco:2014cwa,Blum:2014tka,Bazavov:2018omf}. Assuming $C_{A,P}\sim 1/\Lambda_{A,P}^2$, the sensitivity of these limits translates to mass scales $\Lambda_A\sim 0.1\TeV$, $\Lambda_P\sim 4\TeV$, reflecting the enhancement by $\mpii/(2\sqrt{m_e\hat m})\sim 50$, although the latter is a scale-dependent statement. Matching onto four-fermion operators in SM effective field theory~\cite{Grzadkowski:2010es,Buchmuller:1985jz}, Eq.~\eqref{CACP}
provides constraints for
\begin{align}
 C_A&=\frac{1}{4}\Big(C_{eu}-C_{ed}-C_{\ell u}+C_{\ell d}-2C_{\ell q}^{(3)}\Big),\notag\\
 C_P&=\frac{1}{4}\Big(C_{\ell e q u}^{(1)}-C_{\ell e d q}\Big).
\end{align}
While the one for $C_A$ is not very stringent, the combination of Wilson coefficients differs from the ones probed in parity-violating electron scattering or atomic parity violation, in such a way that the resulting constraint may still be useful to close flat directions in the parameter space, see, e.g., Refs.~\cite{deBlas:2013qqa,Falkowski:2017pss,Crivellin:2021bkd}.
 
Other BSM scenarios include light axial-vector ($Z'$) or pseudoscalar ($a$) states, with minimal couplings
\beq
\Lagr^{(2)}_\text{BSM}=\sum_{f=e,u,d}\bar f\Big(c_A^f \gamma^\mu\gamma_5 Z_\mu'+c_P^f i\gamma_5 a\Big)f,
\eeq
leading to
\beq
\label{Zprime_matching}
C_A=-\frac{(c_A^u-c_A^d)c_A^e}{M_{Z'}^2},\quad 
C_P=\frac{(c_P^u-c_P^d)c_P^e}{m_a^2-q^2},
\eeq
where the new particles correspond to a $Z'$ or an axion-like particle $a$, 
respectively. Writing the $Z'$ interactions in a gauge-invariant way in general 
requires the introduction of Goldstone modes, so that in 
Eq.~\eqref{Zprime_matching} and below we use unitary gauge to make the particle 
content explicit. In this way, the pole in $C_A$ cancels and the SM $Z$ 
contribution is again recovered for $c_A^u=-c_A^d=-c_A^e=g/(4\cos\theta_W)$. In 
contrast, the pseudoscalar pole remains, and the constraints from $\pi^0\to 
e^+e^-$ then also depend on the $\pi^0$--$a$ 
mixing.

As an application, we consider the interplay with the anomalous magnetic moment of the electron $a_e$, which is timely given the current tensions between the direct measurement~\cite{Hanneke:2008tm} and the SM prediction~\cite{Aoyama:2019ryr,Aoyama:2020ynm} either based on the fine-structure constant measured with Cs~\cite{Parker:2018vye} or Rb~\cite{Morel:2020dww} atom interferometry
\begin{align}
\Delta a_e[\text{Cs}]&=a_e^\text{exp}-a_e^\text{SM}[\text{Cs}]=-0.88(36)\times 10^{-12},\notag\\
\Delta a_e[\text{Rb}]&=a_e^\text{exp}-a_e^\text{SM}[\text{Rb}]=0.48(30)\times 10^{-12},
\end{align}
 corresponding to a tension of $-2.5\sigma$ and $+1.6\sigma$, respectively. With the $5.4\sigma$ disagreement between Refs.~\cite{Parker:2018vye,Morel:2020dww} unresolved, we will concentrate here on the case of $\Delta a_e[\text{Cs}]$, since a negative effect can be explained by axial-vector or pseudoscalar mediators~\cite{Leveille:1977rc,Liu:2018xkx}
 \begin{align}
  a_e^A&=-\frac{(c_A^e)^2m_e^2}{4\pi^2M_{Z'}^2}\int_0^1\diff x\frac{2x^3m_e^2+x(1-x)(4-x)M_{Z'}^2}{m_e^2 x^2+M_{Z'}^2(1-x)},\notag\\
  a_e^P&=-\frac{(c_P^e)^2m_e^2}{8\pi^2}\int_0^1\diff x\frac{x^3}{m_e^2 x^2+m_a^2(1-x)},
 \end{align}
while, at one-loop level, vector and scalar mediators yield a positive contribution. For the axial-vector case, the contour plot revises the previously preferred region~\cite{Kahn:2016vjr,Parker:2018vye} according to our improved SM prediction and the radiative corrections~\cite{Vasko:2011pi,Husek:2014tna} applied to the KTeV measurement, see Fig.~\ref{fig:bands}. The parameter regions favored by $\Delta a_e[\text{Cs}]$ and $\pi^0\to e^+e^-$ partly overlap, in which case the quark couplings $c_{A,P}^u-c_{A,P}^d$ take similar values as the electron ones. Note that in Refs.~\cite{Kahn:2016vjr,Parker:2018vye} specific values for the quark couplings have been assumed to show the constraints solely on $c_A^e$; see these references for other constraints on axial-vector $Z'$ models, e.g., from $e^+e^-$ colliders~\cite{Lees:2017lec}. We have further restricted the masses to the parameter region below $\mpii$, and neglected the potential $\pi^0$--$a$ mixing (as before, the pseudoscalar couplings are evaluated at the $\overline{\text{MS}}$ scale $\mu=2\GeV$). Figure~\ref{fig:bands} shows that if both mild tensions were confirmed at this level, similar regions in parameter space seem to be preferred.

\section{Conclusions}

In this Letter we presented an improved SM prediction for the $\pi^0\to e^+e^-$ decay, based on a dispersive representation of the pion transition form factor. This representation---which combines constraints from $\pi^0\to\gamma\gamma$, the low-energy singularities via $e^+e^-\to 2\pi,3\pi$, space-like data for large $Q^2$, and short-distance constraints---allows for a reliable evaluation of the long-range $\gamma^*\gamma^*$ contribution, leading to a SM prediction~\eqref{SM} with a precision of $0.5\%$. The loop integral can be reduced to standard loop functions by means of a double-spectral representation, for which also a Wick rotation to space-like momenta becomes possible. The conceptual advances presented here will also become relevant for refined predictions of the dilepton decays of $\eta^{(\prime)}$.

By comparing our SM prediction to the KTeV measurement we then provided the corresponding constraints on axial-vector and pseudoscalar operators, both in SM effective field theory and for light mediators. In the latter case, we compared the (mildly) favored parameter space to the one suggested by the anomalous magnetic moment of the electron when contrasted to the fine-structure constant measured with Cs atom interferometry. With our calculation of the $\pi^0\to e^+e^-$ width, the theoretical precision now exceeds experiment by an order of magnitude, allowing for concurrent advances in BSM constraints once an improved measurement becomes available. Such efforts are in progress at NA62~\cite{NA62:2713499}.

\begin{acknowledgments}
We thank Tom\'a\v s Husek, Karol Kampf, and Ji\v r\'i Novotn\'y for valuable communication regarding Refs.~\cite{Vasko:2011pi,Husek:2014tna}. 
Financial support by the SNSF (Project No.\ PCEFP2\_181117) and
the DFG through the funds provided to the Sino--German Collaborative
Research Center TRR110 ``Symmetries and the Emergence of Structure in QCD'' (DFG Project-ID 196253076 -- TRR 110)
is gratefully acknowledged.
 J.L.\ is supported by the FWF-DACH Grant I 3845-N27 and by the FWF doctoral program Particles and Interactions, project No.\ W 1252-N27.
\end{acknowledgments}

\appendix

\begin{widetext}

\section{Loop functions}
\label{app:loop_functions}

For the calculation of $\A(q^2)$ using a double-spectral representation of the TFF it suffices to consider the case
\beq
\tilde F_{\pi^0\gamma^*\gamma^*}(q_1^2,q_2^2)=\tilde F_{\pi^0\gamma^*\gamma^*}^\text{VMD}(q_1^2,q_2^2)\equiv\frac{x y}{(x-q_1^2)(y-q_2^2)}
\eeq
for arbitrary $x$, $y$, with the final result obtained by an integration over the corresponding doubly-spectral density. In this case, the decomposition into standard loop functions becomes  
\begin{align}
\A^\text{VMD}(q^2)&=\frac{1}{2q^2}\Big(x \bar B_0(y,m_e)+y \bar B_0(x,m_e)\Big)+L(x,y)-L(x,0)-L(0,y)+L(0,0),\notag\\
L(x,y)&=\frac{\lambda(x,y,q^2)}{2q^2}C_0(q^2,m_e,x,y),
\end{align}
where $\lambda(x,y,z)=x^2+y^2+z^2-2(x y+x z+y z)$ and
\begin{align}
\label{loop_functions}
 \bar B_0(x,m_e)&=\frac{1}{i\pi^2}\int\frac{\diff^4 k}{(k^2-x)\big((p-k)^2-m_e^2\big)}-(x\to 0)\notag\\
 &=-\int_0^1\diff u\log\bigg[1+\frac{x}{m_e^2}\frac{1-u}{u^2}\bigg]
 =\frac{x}{2m_e^2}\bigg[\log\frac{m_e^2}{x}-\sigma_e(x)\log \big[y_e(x)\big]\bigg],\notag\\
 C_0(q^2,m_e,x,y)&=\frac{1}{i\pi^2}\int\frac{\diff^4 k}{(k^2-x)\big((q-k)^2-y\big)\big((p-k)^2-m_e^2\big)}
 =-\int_0^1\diff u\int_0^{1-u}\diff v\, \big[\Delta(x,y,u,v)\big]^{-1},\notag\\
 \Delta(x,y,u,v)&=u x+v y-u vq^2+(1-u-v)^2m_e^2.
\end{align}
In principle, all loop functions can be given in semi-analytic form in terms of polylogarithms~\cite{tHooft:1978jhc}, but for our application the numerical stability in all corners of the double-spectral integration is critical, for which in many cases these representations are not necessarily the most viable ones. We used the implementation from \textsc{LoopTools}~\cite{Hahn:1998yk}, and verified the result using alternative representations based on dispersion relations, Feynman parameterizations, and asymptotic expansions, e.g., the asymptotic contribution can be expressed as
\begin{align}
 \A^\text{asym}(q^2)&=-\frac{2\Fpi}{F_{\pi\gamma\gamma}}\int_{\sm}^\infty \diff x
 \int_0^1\diff u\int_0^{1-u}\diff v\, uv\bigg[\frac{3}{[\Delta(x,x,u,v)]^2}+\frac{(1-u-v)^2\big(q^2-4m_e^2\big)}{[\Delta(x,x,u,v)]^3}\bigg].
\end{align}
A potential alternative strategy could proceed via a Wick rotation to the space-like region and subsequent Gegenbauer integration of 
the angular integrals~\cite{Rosner:1967zz,Levine:1974xh,Levine:1979uz}. For a general TFF such a strategy cannot work due to the dependence on both $k^2$ and $(q-k)^2$, but once a double-spectral representation is employed, a Wick rotation becomes possible when first combining the two photon propagators with a Feynman parameter $u$ and then performing the Gegenbauer average with respect to the shifted momentum $p-uq$. The result for the loop functions becomes
\begin{align}
 \bar B_0(x,m_e)&=\frac{x}{m_e^2}\int_0^\infty\frac{\diff K\,K}{K^2+x}\bigg[1-\sqrt{1+\frac{4m_e^2}{K^2}}\bigg],\notag\\
  C_0(q^2,m_e,x,y)&=\int_0^1\diff u\int_0^\infty\diff K\, K\frac{K^2+u(1-u)q^2-\sqrt{4K^2m_e^2+\big(K^2-u(1-u)q^2\big)^2}}{[K^2+u x+(1-u) y-u(1-u)q^2]^2(m_e^2-u(1-u)q^2)}.
\end{align}
The numerical stability can be improved by a change of variables $K\to K\sqrt{u(1-u)q^2}$ and by separating the imaginary part that arises for $x=y=0$, but for the evaluation of the double-spectral integral the result is still not viable. 

However, the fact that it is possible to rotate the loop integral over the TFF to space-like kinematics suggests that this region ultimately determines the decay amplitude. To verify that the matching of the three contributions to the TFF indeed applies as expected, we thus consider a similar representation for the TFF directly 
\begin{align}
\label{Fspace}
\tilde F_{\pi^0\gamma^*\gamma^*}\big(k^2,(q-k)^2\big)&=\int_0^1\diff u\bigg[\frac{1}{\pi^2}\int\diff x\int\diff y \frac{\tilde \rho(x,y)}{[\tilde k^2-\Delta(x,y,u)]^2}
+g_\text{eff}\frac{M_\text{eff}^4}{[\tilde k^2-\Delta(M_\text{eff}^2,u)]^2}\notag\\
&+\frac{2F_\pi}{F_{\pi\gamma\gamma}}\int_{\sm}^\infty\diff x\bigg(\frac{1}{[\tilde k^2-\Delta(x,u)]^2}+\frac{6x^2u(1-u)}{[\tilde k^2-\Delta(x,u)]^4}+\frac{2x}{[\tilde k^2-\Delta(x,u)]^3}\bigg)\bigg],
\end{align}
where
\begin{align}
\Delta(x,y,u)&=x(1-u)+y u-u(1-u)q^2,\qquad \Delta(x,u)=x-u(1-u)q^2,\qquad \tilde k=k-uq,
\end{align}
and, after Wick rotation, 
$\tilde k^2\to - Q^2$ could be interpreted as a space-like momentum. The resulting form factor $\tilde F(-Q^2)$ has been used in Fig.~\ref{fig:TFF} to illustrate the different contributions to the pion TFF, i.e.,  $\tilde F(-Q^2)$ is defined by Eq.~\eqref{Fspace} with $\tilde k^2=-Q^2$ and $q^2=\mpii^2$ (see Fig.~\ref{fig:TFF_low} for a variant focused on the low-energy region). 

\begin{figure}[t]
	\includegraphics[width=0.5\linewidth]{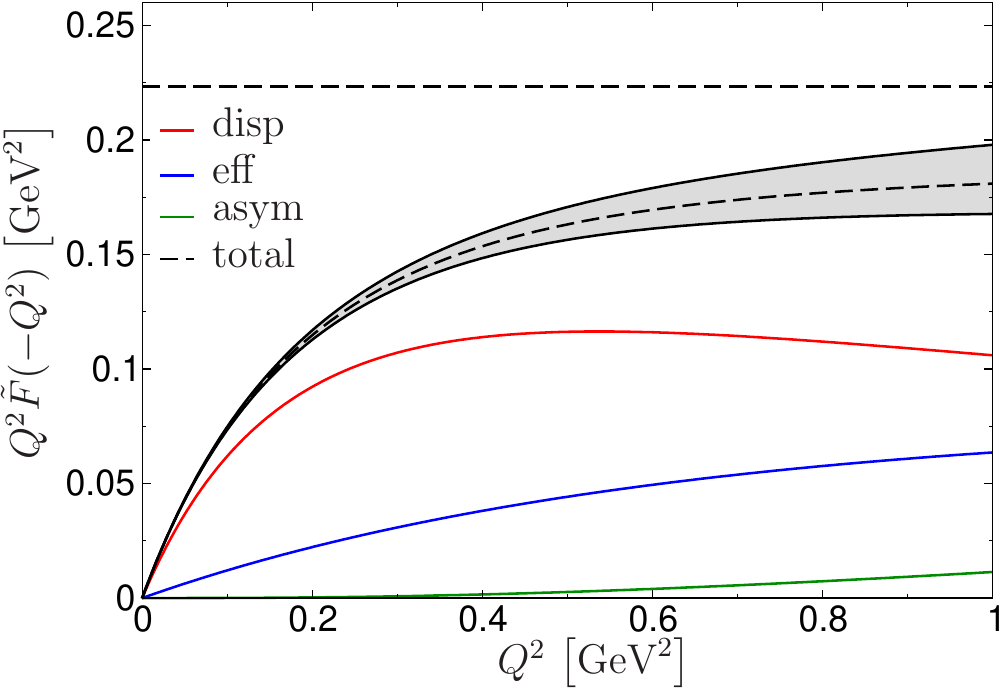}
	\caption{Same as Fig.~\ref{fig:TFF}, but focused on the low-energy region.} 
	\label{fig:TFF_low}
\end{figure}

In the limit $q^2\to 0$, Eq.~\eqref{Fspace} then simply reduces to the diagonal doubly-virtual TFF $\tilde F_{\pi^0\gamma^*\gamma^*}\big(-Q^2,-Q^2\big)$. Given that $q^2\ll M_\rho^2$, combined with the additional suppression from the $u$ integral, this explains why an approximate formula based on a dispersion relation in $q^2$ (and the diagonal doubly-virtual TFF)~\cite{Dorokhov:2007bd,Dorokhov:2008cd} 
\begin{align}
\label{Aq2}
 \Re\A(q^2)&=\A(0)+\frac{1}{\sigma_e(q^2)}\bigg[\text{Li}_2\big[-y_e(q^2)\big]
 +\frac{1}{4}\log^2\big[y_e(q^2)]+\frac{\pi^2}{12}\bigg],\notag\\
 \A(0)&\approx 3\log\frac{m_e}{\mu}-\frac{3}{2}\bigg[\int_0^{\mu^2}\diff t\frac{\tilde F_{\pi^0\gamma^*\gamma^*}(-t,-t)-1}{t}
 +\int_{\mu^2}^\infty \diff t\frac{\tilde F_{\pi^0\gamma^*\gamma^*}(-t,-t)}{t}\bigg]-\frac{5}{4},
\end{align}
is close to the full result from the double-spectral representation:
\beq
\Re \A(q^2)|_{\gamma^*\gamma^*}\approx 10.00(4)_\text{disp}(8)_\text{BL}(2)_\text{asym},
\eeq
with the relative difference to Eq.~\eqref{disp_result}  about $1.5\%$ compared to $q^2/M_\rho^2\approx 3\%$. However, we stress that apart from mass expansions Eq.~\eqref{Aq2} involves an unquantified uncertainty because the TFF is not an observable away from $q^2=\mpii^2$. It is also restricted to the $\gamma\gamma$ cut, so that the corresponding approximation becomes worse for heavier pseudoscalars. 
\end{widetext}

\bibliography{pi0ee}

\begin{thebibliography}{117}
\expandafter\ifx\csname natexlab\endcsname\relax\def\natexlab#1{#1}\fi
\expandafter\ifx\csname bibnamefont\endcsname\relax
  \def\bibnamefont#1{#1}\fi
\expandafter\ifx\csname bibfnamefont\endcsname\relax
  \def\bibfnamefont#1{#1}\fi
\expandafter\ifx\csname citenamefont\endcsname\relax
  \def\citenamefont#1{#1}\fi
\expandafter\ifx\csname url\endcsname\relax
  \def\url#1{\texttt{#1}}\fi
\expandafter\ifx\csname urlprefix\endcsname\relax\def\urlprefix{URL }\fi
\providecommand{\bibinfo}[2]{#2}
\providecommand{\eprint}[2][]{\url{#2}}

\bibitem[{\citenamefont{Wess and Zumino}(1971)}]{Wess:1971yu}
\bibinfo{author}{\bibfnamefont{J.}~\bibnamefont{Wess}} \bibnamefont{and}
  \bibinfo{author}{\bibfnamefont{B.}~\bibnamefont{Zumino}},
  \bibinfo{journal}{Phys. Lett. B} \textbf{\bibinfo{volume}{37}},
  \bibinfo{pages}{95} (\bibinfo{year}{1971}).

\bibitem[{\citenamefont{Witten}(1983)}]{Witten:1983tw}
\bibinfo{author}{\bibfnamefont{E.}~\bibnamefont{Witten}},
  \bibinfo{journal}{Nucl. Phys. B} \textbf{\bibinfo{volume}{223}},
  \bibinfo{pages}{422} (\bibinfo{year}{1983}).

\bibitem[{\citenamefont{Adler}(1969)}]{Adler:1969gk}
\bibinfo{author}{\bibfnamefont{S.~L.} \bibnamefont{Adler}},
  \bibinfo{journal}{Phys. Rev.} \textbf{\bibinfo{volume}{177}},
  \bibinfo{pages}{2426} (\bibinfo{year}{1969}).

\bibitem[{\citenamefont{Bell and Jackiw}(1969)}]{Bell:1969ts}
\bibinfo{author}{\bibfnamefont{J.~S.} \bibnamefont{Bell}} \bibnamefont{and}
  \bibinfo{author}{\bibfnamefont{R.}~\bibnamefont{Jackiw}},
  \bibinfo{journal}{Nuovo Cim. A} \textbf{\bibinfo{volume}{60}},
  \bibinfo{pages}{47} (\bibinfo{year}{1969}).

\bibitem[{\citenamefont{Bardeen}(1969)}]{Bardeen:1969md}
\bibinfo{author}{\bibfnamefont{W.~A.} \bibnamefont{Bardeen}},
  \bibinfo{journal}{Phys. Rev.} \textbf{\bibinfo{volume}{184}},
  \bibinfo{pages}{1848} (\bibinfo{year}{1969}).

\bibitem[{\citenamefont{Zyla et~al.}(2020)}]{Zyla:2020zbs}
\bibinfo{author}{\bibfnamefont{P.~A.} \bibnamefont{Zyla}} \bibnamefont{et~al.}
  (\bibinfo{collaboration}{Particle Data Group}), \bibinfo{journal}{PTEP}
  \textbf{\bibinfo{volume}{2020}}, \bibinfo{pages}{083C01}
  (\bibinfo{year}{2020}).

\bibitem[{\citenamefont{Larin et~al.}(2020)}]{Larin:2020bhc}
\bibinfo{author}{\bibfnamefont{I.}~\bibnamefont{Larin}} \bibnamefont{et~al.}
  (\bibinfo{collaboration}{PrimEx-II}), \bibinfo{journal}{Science}
  \textbf{\bibinfo{volume}{368}}, \bibinfo{pages}{506} (\bibinfo{year}{2020}).

\bibitem[{\citenamefont{Bijnens et~al.}(1990)\citenamefont{Bijnens, Bramon, and
  Cornet}}]{Bijnens:1989jb}
\bibinfo{author}{\bibfnamefont{J.}~\bibnamefont{Bijnens}},
  \bibinfo{author}{\bibfnamefont{A.}~\bibnamefont{Bramon}}, \bibnamefont{and}
  \bibinfo{author}{\bibfnamefont{F.}~\bibnamefont{Cornet}},
  \bibinfo{journal}{Z. Phys. C} \textbf{\bibinfo{volume}{46}},
  \bibinfo{pages}{599} (\bibinfo{year}{1990}).

\bibitem[{\citenamefont{Ananthanarayan and
  Moussallam}(2002)}]{Ananthanarayan:2002kj}
\bibinfo{author}{\bibfnamefont{B.}~\bibnamefont{Ananthanarayan}}
  \bibnamefont{and}
  \bibinfo{author}{\bibfnamefont{B.}~\bibnamefont{Moussallam}},
  \bibinfo{journal}{JHEP} \textbf{\bibinfo{volume}{05}}, \bibinfo{pages}{052}
  (\bibinfo{year}{2002}), \eprint{hep-ph/0205232}.

\bibitem[{\citenamefont{Goity et~al.}(2002)\citenamefont{Goity, Bernstein, and
  Holstein}}]{Goity:2002nn}
\bibinfo{author}{\bibfnamefont{J.~L.} \bibnamefont{Goity}},
  \bibinfo{author}{\bibfnamefont{A.~M.} \bibnamefont{Bernstein}},
  \bibnamefont{and} \bibinfo{author}{\bibfnamefont{B.~R.}
  \bibnamefont{Holstein}}, \bibinfo{journal}{Phys. Rev. D}
  \textbf{\bibinfo{volume}{66}}, \bibinfo{pages}{076014}
  (\bibinfo{year}{2002}), \eprint{hep-ph/0206007}.

\bibitem[{\citenamefont{Kampf and Moussallam}(2009)}]{Kampf:2009tk}
\bibinfo{author}{\bibfnamefont{K.}~\bibnamefont{Kampf}} \bibnamefont{and}
  \bibinfo{author}{\bibfnamefont{B.}~\bibnamefont{Moussallam}},
  \bibinfo{journal}{Phys. Rev. D} \textbf{\bibinfo{volume}{79}},
  \bibinfo{pages}{076005} (\bibinfo{year}{2009}), \eprint{0901.4688}.

\bibitem[{\citenamefont{Husek et~al.}(2015)\citenamefont{Husek, Kampf, and
  Novotn\'y}}]{Husek:2015sma}
\bibinfo{author}{\bibfnamefont{T.}~\bibnamefont{Husek}},
  \bibinfo{author}{\bibfnamefont{K.}~\bibnamefont{Kampf}}, \bibnamefont{and}
  \bibinfo{author}{\bibfnamefont{J.}~\bibnamefont{Novotn\'y}},
  \bibinfo{journal}{Phys. Rev. D} \textbf{\bibinfo{volume}{92}},
  \bibinfo{pages}{054027} (\bibinfo{year}{2015}), \eprint{1504.06178}.

\bibitem[{\citenamefont{Hoferichter
  et~al.}(2014{\natexlab{a}})\citenamefont{Hoferichter, Kubis, Leupold,
  Niecknig, and Schneider}}]{Hoferichter:2014vra}
\bibinfo{author}{\bibfnamefont{M.}~\bibnamefont{Hoferichter}},
  \bibinfo{author}{\bibfnamefont{B.}~\bibnamefont{Kubis}},
  \bibinfo{author}{\bibfnamefont{S.}~\bibnamefont{Leupold}},
  \bibinfo{author}{\bibfnamefont{F.}~\bibnamefont{Niecknig}}, \bibnamefont{and}
  \bibinfo{author}{\bibfnamefont{S.~P.} \bibnamefont{Schneider}},
  \bibinfo{journal}{Eur. Phys. J. C} \textbf{\bibinfo{volume}{74}},
  \bibinfo{pages}{3180} (\bibinfo{year}{2014}{\natexlab{a}}),
  \eprint{1410.4691}.

\bibitem[{\citenamefont{Hoferichter
  et~al.}(2018{\natexlab{a}})\citenamefont{Hoferichter, Hoid, Kubis, Leupold,
  and Schneider}}]{Hoferichter:2018dmo}
\bibinfo{author}{\bibfnamefont{M.}~\bibnamefont{Hoferichter}},
  \bibinfo{author}{\bibfnamefont{B.-L.} \bibnamefont{Hoid}},
  \bibinfo{author}{\bibfnamefont{B.}~\bibnamefont{Kubis}},
  \bibinfo{author}{\bibfnamefont{S.}~\bibnamefont{Leupold}}, \bibnamefont{and}
  \bibinfo{author}{\bibfnamefont{S.~P.} \bibnamefont{Schneider}},
  \bibinfo{journal}{Phys. Rev. Lett.} \textbf{\bibinfo{volume}{121}},
  \bibinfo{pages}{112002} (\bibinfo{year}{2018}{\natexlab{a}}),
  \eprint{1805.01471}.

\bibitem[{\citenamefont{Hoferichter
  et~al.}(2018{\natexlab{b}})\citenamefont{Hoferichter, Hoid, Kubis, Leupold,
  and Schneider}}]{Hoferichter:2018kwz}
\bibinfo{author}{\bibfnamefont{M.}~\bibnamefont{Hoferichter}},
  \bibinfo{author}{\bibfnamefont{B.-L.} \bibnamefont{Hoid}},
  \bibinfo{author}{\bibfnamefont{B.}~\bibnamefont{Kubis}},
  \bibinfo{author}{\bibfnamefont{S.}~\bibnamefont{Leupold}}, \bibnamefont{and}
  \bibinfo{author}{\bibfnamefont{S.~P.} \bibnamefont{Schneider}},
  \bibinfo{journal}{JHEP} \textbf{\bibinfo{volume}{10}}, \bibinfo{pages}{141}
  (\bibinfo{year}{2018}{\natexlab{b}}), \eprint{1808.04823}.

\bibitem[{\citenamefont{Masjuan}(2012)}]{Masjuan:2012wy}
\bibinfo{author}{\bibfnamefont{P.}~\bibnamefont{Masjuan}},
  \bibinfo{journal}{Phys. Rev. D} \textbf{\bibinfo{volume}{86}},
  \bibinfo{pages}{094021} (\bibinfo{year}{2012}), \eprint{1206.2549}.

\bibitem[{\citenamefont{Masjuan and S\'anchez-Puertas}(2017)}]{Masjuan:2017tvw}
\bibinfo{author}{\bibfnamefont{P.}~\bibnamefont{Masjuan}} \bibnamefont{and}
  \bibinfo{author}{\bibfnamefont{P.}~\bibnamefont{S\'anchez-Puertas}},
  \bibinfo{journal}{Phys. Rev. D} \textbf{\bibinfo{volume}{95}},
  \bibinfo{pages}{054026} (\bibinfo{year}{2017}), \eprint{1701.05829}.

\bibitem[{\citenamefont{Behrend et~al.}(1991)}]{Behrend:1990sr}
\bibinfo{author}{\bibfnamefont{H.~J.} \bibnamefont{Behrend}}
  \bibnamefont{et~al.} (\bibinfo{collaboration}{CELLO}), \bibinfo{journal}{Z.
  Phys. C} \textbf{\bibinfo{volume}{49}}, \bibinfo{pages}{401}
  (\bibinfo{year}{1991}).

\bibitem[{\citenamefont{Adlarson et~al.}(2017)}]{Adlarson:2016ykr}
\bibinfo{author}{\bibfnamefont{P.}~\bibnamefont{Adlarson}} \bibnamefont{et~al.}
  (\bibinfo{collaboration}{A2}), \bibinfo{journal}{Phys. Rev. C}
  \textbf{\bibinfo{volume}{95}}, \bibinfo{pages}{025202}
  (\bibinfo{year}{2017}), \eprint{1611.04739}.

\bibitem[{\citenamefont{Lazzeroni et~al.}(2017)}]{TheNA62:2016fhr}
\bibinfo{author}{\bibfnamefont{C.}~\bibnamefont{Lazzeroni}}
  \bibnamefont{et~al.} (\bibinfo{collaboration}{NA62}), \bibinfo{journal}{Phys.
  Lett. B} \textbf{\bibinfo{volume}{768}}, \bibinfo{pages}{38}
  (\bibinfo{year}{2017}), \eprint{1612.08162}.

\bibitem[{\citenamefont{Husek et~al.}(2019)\citenamefont{Husek, Goudzovski, and
  Kampf}}]{Husek:2018qdx}
\bibinfo{author}{\bibfnamefont{T.}~\bibnamefont{Husek}},
  \bibinfo{author}{\bibfnamefont{E.}~\bibnamefont{Goudzovski}},
  \bibnamefont{and} \bibinfo{author}{\bibfnamefont{K.}~\bibnamefont{Kampf}},
  \bibinfo{journal}{Phys. Rev. Lett.} \textbf{\bibinfo{volume}{122}},
  \bibinfo{pages}{022003} (\bibinfo{year}{2019}), \eprint{1809.01153}.

\bibitem[{\citenamefont{Samios}(1961)}]{Samios:1961zz}
\bibinfo{author}{\bibfnamefont{N.~P.} \bibnamefont{Samios}},
  \bibinfo{journal}{Phys. Rev.} \textbf{\bibinfo{volume}{121}},
  \bibinfo{pages}{275} (\bibinfo{year}{1961}).

\bibitem[{\citenamefont{Schardt et~al.}(1981)\citenamefont{Schardt, Frank,
  Hoffman, Mischke, Moir, and Thompson}}]{Schardt:1980qd}
\bibinfo{author}{\bibfnamefont{M.~A.} \bibnamefont{Schardt}},
  \bibinfo{author}{\bibfnamefont{J.~S.} \bibnamefont{Frank}},
  \bibinfo{author}{\bibfnamefont{C.~M.} \bibnamefont{Hoffman}},
  \bibinfo{author}{\bibfnamefont{R.~E.} \bibnamefont{Mischke}},
  \bibinfo{author}{\bibfnamefont{D.~C.} \bibnamefont{Moir}}, \bibnamefont{and}
  \bibinfo{author}{\bibfnamefont{P.~A.} \bibnamefont{Thompson}},
  \bibinfo{journal}{Phys. Rev. D} \textbf{\bibinfo{volume}{23}},
  \bibinfo{pages}{639} (\bibinfo{year}{1981}).

\bibitem[{\citenamefont{Beddall}(2008)}]{Beddall:2008zza}
\bibinfo{author}{\bibfnamefont{A.}~\bibnamefont{Beddall}},
  \bibinfo{journal}{Eur. Phys. J. C} \textbf{\bibinfo{volume}{54}},
  \bibinfo{pages}{365} (\bibinfo{year}{2008}).

\bibitem[{\citenamefont{Abouzaid et~al.}(2008)}]{Abouzaid:2008cd}
\bibinfo{author}{\bibfnamefont{E.}~\bibnamefont{Abouzaid}} \bibnamefont{et~al.}
  (\bibinfo{collaboration}{KTeV}), \bibinfo{journal}{Phys. Rev. Lett.}
  \textbf{\bibinfo{volume}{100}}, \bibinfo{pages}{182001}
  (\bibinfo{year}{2008}), \eprint{0802.2064}.

\bibitem[{\citenamefont{Drell}(1959)}]{Drell:1959}
\bibinfo{author}{\bibfnamefont{S.~D.} \bibnamefont{Drell}},
  \bibinfo{journal}{Nuovo Cim.} \textbf{\bibinfo{volume}{11}},
  \bibinfo{pages}{693} (\bibinfo{year}{1959}).

\bibitem[{\citenamefont{Soni}(1974)}]{Soni:1974aw}
\bibinfo{author}{\bibfnamefont{A.}~\bibnamefont{Soni}}, \bibinfo{journal}{Phys.
  Lett. B} \textbf{\bibinfo{volume}{52}}, \bibinfo{pages}{332}
  (\bibinfo{year}{1974}).

\bibitem[{\citenamefont{Arnellos et~al.}(1982)\citenamefont{Arnellos, Marciano,
  and Parsa}}]{Arnellos:1981bk}
\bibinfo{author}{\bibfnamefont{L.}~\bibnamefont{Arnellos}},
  \bibinfo{author}{\bibfnamefont{W.~J.} \bibnamefont{Marciano}},
  \bibnamefont{and} \bibinfo{author}{\bibfnamefont{Z.}~\bibnamefont{Parsa}},
  \bibinfo{journal}{Nucl. Phys. B} \textbf{\bibinfo{volume}{196}},
  \bibinfo{pages}{365} (\bibinfo{year}{1982}).

\bibitem[{\citenamefont{Masjuan and S\'anchez-Puertas}(2015)}]{Masjuan:2015lca}
\bibinfo{author}{\bibfnamefont{P.}~\bibnamefont{Masjuan}} \bibnamefont{and}
  \bibinfo{author}{\bibfnamefont{P.}~\bibnamefont{S\'anchez-Puertas}}
  (\bibinfo{year}{2015}), \eprint{1504.07001}.

\bibitem[{\citenamefont{Masjuan and S\'anchez-Puertas}(2016)}]{Masjuan:2015cjl}
\bibinfo{author}{\bibfnamefont{P.}~\bibnamefont{Masjuan}} \bibnamefont{and}
  \bibinfo{author}{\bibfnamefont{P.}~\bibnamefont{S\'anchez-Puertas}},
  \bibinfo{journal}{JHEP} \textbf{\bibinfo{volume}{08}}, \bibinfo{pages}{108}
  (\bibinfo{year}{2016}), \eprint{1512.09292}.

\bibitem[{\citenamefont{Kahn et~al.}(2008)\citenamefont{Kahn, Schmitt, and
  Tait}}]{Kahn:2007ru}
\bibinfo{author}{\bibfnamefont{Y.}~\bibnamefont{Kahn}},
  \bibinfo{author}{\bibfnamefont{M.}~\bibnamefont{Schmitt}}, \bibnamefont{and}
  \bibinfo{author}{\bibfnamefont{T.~M.~P.} \bibnamefont{Tait}},
  \bibinfo{journal}{Phys. Rev. D} \textbf{\bibinfo{volume}{78}},
  \bibinfo{pages}{115002} (\bibinfo{year}{2008}), \eprint{0712.0007}.

\bibitem[{\citenamefont{Kahn et~al.}(2017)\citenamefont{Kahn, Krnjaic,
  Mishra-Sharma, and Tait}}]{Kahn:2016vjr}
\bibinfo{author}{\bibfnamefont{Y.}~\bibnamefont{Kahn}},
  \bibinfo{author}{\bibfnamefont{G.}~\bibnamefont{Krnjaic}},
  \bibinfo{author}{\bibfnamefont{S.}~\bibnamefont{Mishra-Sharma}},
  \bibnamefont{and} \bibinfo{author}{\bibfnamefont{T.~M.~P.}
  \bibnamefont{Tait}}, \bibinfo{journal}{JHEP} \textbf{\bibinfo{volume}{05}},
  \bibinfo{pages}{002} (\bibinfo{year}{2017}), \eprint{1609.09072}.

\bibitem[{\citenamefont{Chang and Yang}(2009)}]{Chang:2008np}
\bibinfo{author}{\bibfnamefont{Q.}~\bibnamefont{Chang}} \bibnamefont{and}
  \bibinfo{author}{\bibfnamefont{Y.-D.} \bibnamefont{Yang}},
  \bibinfo{journal}{Phys. Lett. B} \textbf{\bibinfo{volume}{676}},
  \bibinfo{pages}{88} (\bibinfo{year}{2009}), \eprint{0808.2933}.

\bibitem[{\citenamefont{Andreas et~al.}(2010)\citenamefont{Andreas, Lebedev,
  Ramos-S\'anchez, and Ringwald}}]{Andreas:2010ms}
\bibinfo{author}{\bibfnamefont{S.}~\bibnamefont{Andreas}},
  \bibinfo{author}{\bibfnamefont{O.}~\bibnamefont{Lebedev}},
  \bibinfo{author}{\bibfnamefont{S.}~\bibnamefont{Ramos-S\'anchez}},
  \bibnamefont{and} \bibinfo{author}{\bibfnamefont{A.}~\bibnamefont{Ringwald}},
  \bibinfo{journal}{JHEP} \textbf{\bibinfo{volume}{08}}, \bibinfo{pages}{003}
  (\bibinfo{year}{2010}), \eprint{1005.3978}.

\bibitem[{\citenamefont{Bauer et~al.}(2017)\citenamefont{Bauer, Neubert, and
  Thamm}}]{Bauer:2017ris}
\bibinfo{author}{\bibfnamefont{M.}~\bibnamefont{Bauer}},
  \bibinfo{author}{\bibfnamefont{M.}~\bibnamefont{Neubert}}, \bibnamefont{and}
  \bibinfo{author}{\bibfnamefont{A.}~\bibnamefont{Thamm}},
  \bibinfo{journal}{JHEP} \textbf{\bibinfo{volume}{12}}, \bibinfo{pages}{044}
  (\bibinfo{year}{2017}), \eprint{1708.00443}.

\bibitem[{\citenamefont{Alves and Weiner}(2018)}]{Alves:2017avw}
\bibinfo{author}{\bibfnamefont{D.~S.~M.} \bibnamefont{Alves}} \bibnamefont{and}
  \bibinfo{author}{\bibfnamefont{N.}~\bibnamefont{Weiner}},
  \bibinfo{journal}{JHEP} \textbf{\bibinfo{volume}{07}}, \bibinfo{pages}{092}
  (\bibinfo{year}{2018}), \eprint{1710.03764}.

\bibitem[{\citenamefont{Altmannshofer et~al.}(2020)\citenamefont{Altmannshofer,
  Gori, and Robinson}}]{Altmannshofer:2019yji}
\bibinfo{author}{\bibfnamefont{W.}~\bibnamefont{Altmannshofer}},
  \bibinfo{author}{\bibfnamefont{S.}~\bibnamefont{Gori}}, \bibnamefont{and}
  \bibinfo{author}{\bibfnamefont{D.~J.} \bibnamefont{Robinson}},
  \bibinfo{journal}{Phys. Rev. D} \textbf{\bibinfo{volume}{101}},
  \bibinfo{pages}{075002} (\bibinfo{year}{2020}), \eprint{1909.00005}.

\bibitem[{\citenamefont{Bauer et~al.}(2021)\citenamefont{Bauer, Neubert,
  Renner, Schnubel, and Thamm}}]{Bauer:2021mvw}
\bibinfo{author}{\bibfnamefont{M.}~\bibnamefont{Bauer}},
  \bibinfo{author}{\bibfnamefont{M.}~\bibnamefont{Neubert}},
  \bibinfo{author}{\bibfnamefont{S.}~\bibnamefont{Renner}},
  \bibinfo{author}{\bibfnamefont{M.}~\bibnamefont{Schnubel}}, \bibnamefont{and}
  \bibinfo{author}{\bibfnamefont{A.}~\bibnamefont{Thamm}}
  (\bibinfo{year}{2021}), \eprint{2110.10698}.

\bibitem[{\citenamefont{Abouzaid et~al.}(2007)}]{Abouzaid:2006kk}
\bibinfo{author}{\bibfnamefont{E.}~\bibnamefont{Abouzaid}} \bibnamefont{et~al.}
  (\bibinfo{collaboration}{KTeV}), \bibinfo{journal}{Phys. Rev. D}
  \textbf{\bibinfo{volume}{75}}, \bibinfo{pages}{012004}
  (\bibinfo{year}{2007}), \eprint{hep-ex/0610072}.

\bibitem[{\citenamefont{Va{\v s}ko and Novotn\'y}(2011)}]{Vasko:2011pi}
\bibinfo{author}{\bibfnamefont{P.}~\bibnamefont{Va{\v s}ko}} \bibnamefont{and}
  \bibinfo{author}{\bibfnamefont{J.}~\bibnamefont{Novotn\'y}},
  \bibinfo{journal}{JHEP} \textbf{\bibinfo{volume}{10}}, \bibinfo{pages}{122}
  (\bibinfo{year}{2011}), \eprint{1106.5956}.

\bibitem[{\citenamefont{Husek et~al.}(2014)\citenamefont{Husek, Kampf, and
  Novotn\'y}}]{Husek:2014tna}
\bibinfo{author}{\bibfnamefont{T.}~\bibnamefont{Husek}},
  \bibinfo{author}{\bibfnamefont{K.}~\bibnamefont{Kampf}}, \bibnamefont{and}
  \bibinfo{author}{\bibfnamefont{J.}~\bibnamefont{Novotn\'y}},
  \bibinfo{journal}{Eur. Phys. J. C} \textbf{\bibinfo{volume}{74}},
  \bibinfo{pages}{3010} (\bibinfo{year}{2014}), \eprint{1405.6927}.

\bibitem[{\citenamefont{Bergstr{\"o}m}(1983)}]{Bergstrom:1982wk}
\bibinfo{author}{\bibfnamefont{L.}~\bibnamefont{Bergstr{\"o}m}},
  \bibinfo{journal}{Z. Phys. C} \textbf{\bibinfo{volume}{20}},
  \bibinfo{pages}{135} (\bibinfo{year}{1983}).

\bibitem[{\citenamefont{Dorokhov et~al.}(2008)\citenamefont{Dorokhov, Kuraev,
  Bystritskiy, and Secansky}}]{Dorokhov:2008qn}
\bibinfo{author}{\bibfnamefont{A.~E.} \bibnamefont{Dorokhov}},
  \bibinfo{author}{\bibfnamefont{E.~A.} \bibnamefont{Kuraev}},
  \bibinfo{author}{\bibfnamefont{Y.~M.} \bibnamefont{Bystritskiy}},
  \bibnamefont{and} \bibinfo{author}{\bibfnamefont{M.}~\bibnamefont{Secansky}},
  \bibinfo{journal}{Eur. Phys. J. C} \textbf{\bibinfo{volume}{55}},
  \bibinfo{pages}{193} (\bibinfo{year}{2008}), \eprint{0801.2028}.

\bibitem[{\citenamefont{Husek et~al.}(2021)\citenamefont{Husek, Kampf, and
  Novotn\'y}}]{Husek:2021}
\bibinfo{author}{\bibfnamefont{T.}~\bibnamefont{Husek}},
  \bibinfo{author}{\bibfnamefont{K.}~\bibnamefont{Kampf}}, \bibnamefont{and}
  \bibinfo{author}{\bibfnamefont{J.}~\bibnamefont{Novotn\'y}}
  (\bibinfo{year}{2021}), \eprint{private communication}.

\bibitem[{\citenamefont{Berman and Geffen}(1960)}]{Berman:1960zz}
\bibinfo{author}{\bibfnamefont{S.}~\bibnamefont{Berman}} \bibnamefont{and}
  \bibinfo{author}{\bibfnamefont{D.}~\bibnamefont{Geffen}},
  \bibinfo{journal}{Nuovo Cim.} \textbf{\bibinfo{volume}{18}},
  \bibinfo{pages}{1192} (\bibinfo{year}{1960}).

\bibitem[{\citenamefont{Pratap and Smith}(1972)}]{Pratap:1972tb}
\bibinfo{author}{\bibfnamefont{M.}~\bibnamefont{Pratap}} \bibnamefont{and}
  \bibinfo{author}{\bibfnamefont{J.}~\bibnamefont{Smith}},
  \bibinfo{journal}{Phys. Rev. D} \textbf{\bibinfo{volume}{5}},
  \bibinfo{pages}{2020} (\bibinfo{year}{1972}).

\bibitem[{\citenamefont{Gan et~al.}(2022)\citenamefont{Gan, Kubis, Passemar,
  and Tulin}}]{Gan:2020aco}
\bibinfo{author}{\bibfnamefont{L.}~\bibnamefont{Gan}},
  \bibinfo{author}{\bibfnamefont{B.}~\bibnamefont{Kubis}},
  \bibinfo{author}{\bibfnamefont{E.}~\bibnamefont{Passemar}}, \bibnamefont{and}
  \bibinfo{author}{\bibfnamefont{S.}~\bibnamefont{Tulin}},
  \bibinfo{journal}{Phys. Rept.} \textbf{\bibinfo{volume}{945}},
  \bibinfo{pages}{1} (\bibinfo{year}{2022}), \eprint{2007.00664}.

\bibitem[{\citenamefont{Savage et~al.}(1992)\citenamefont{Savage, Luke, and
  Wise}}]{Savage:1992ac}
\bibinfo{author}{\bibfnamefont{M.~J.} \bibnamefont{Savage}},
  \bibinfo{author}{\bibfnamefont{M.~E.} \bibnamefont{Luke}}, \bibnamefont{and}
  \bibinfo{author}{\bibfnamefont{M.~B.} \bibnamefont{Wise}},
  \bibinfo{journal}{Phys. Lett. B} \textbf{\bibinfo{volume}{291}},
  \bibinfo{pages}{481} (\bibinfo{year}{1992}), \eprint{hep-ph/9207233}.

\bibitem[{\citenamefont{G{\'o}mez~Dumm and Pich}(1998)}]{GomezDumm:1998gw}
\bibinfo{author}{\bibfnamefont{D.}~\bibnamefont{G{\'o}mez~Dumm}}
  \bibnamefont{and} \bibinfo{author}{\bibfnamefont{A.}~\bibnamefont{Pich}},
  \bibinfo{journal}{Phys. Rev. Lett.} \textbf{\bibinfo{volume}{80}},
  \bibinfo{pages}{4633} (\bibinfo{year}{1998}), \eprint{hep-ph/9801298}.

\bibitem[{\citenamefont{Ametller et~al.}(1993)\citenamefont{Ametller, Bramon,
  and Mass{\'o}}}]{Ametller:1993we}
\bibinfo{author}{\bibfnamefont{L.}~\bibnamefont{Ametller}},
  \bibinfo{author}{\bibfnamefont{A.}~\bibnamefont{Bramon}}, \bibnamefont{and}
  \bibinfo{author}{\bibfnamefont{E.}~\bibnamefont{Mass{\'o}}},
  \bibinfo{journal}{Phys. Rev. D} \textbf{\bibinfo{volume}{48}},
  \bibinfo{pages}{3388} (\bibinfo{year}{1993}), \eprint{hep-ph/9302304}.

\bibitem[{\citenamefont{Knecht et~al.}(1999)\citenamefont{Knecht, Peris,
  Perrottet, and de~Rafael}}]{Knecht:1999gb}
\bibinfo{author}{\bibfnamefont{M.}~\bibnamefont{Knecht}},
  \bibinfo{author}{\bibfnamefont{S.}~\bibnamefont{Peris}},
  \bibinfo{author}{\bibfnamefont{M.}~\bibnamefont{Perrottet}},
  \bibnamefont{and}
  \bibinfo{author}{\bibfnamefont{E.}~\bibnamefont{de~Rafael}},
  \bibinfo{journal}{Phys. Rev. Lett.} \textbf{\bibinfo{volume}{83}},
  \bibinfo{pages}{5230} (\bibinfo{year}{1999}), \eprint{hep-ph/9908283}.

\bibitem[{\citenamefont{Silagadze}(2006)}]{Silagadze:2006rt}
\bibinfo{author}{\bibfnamefont{Z.~K.} \bibnamefont{Silagadze}},
  \bibinfo{journal}{Phys. Rev. D} \textbf{\bibinfo{volume}{74}},
  \bibinfo{pages}{054003} (\bibinfo{year}{2006}), \eprint{hep-ph/0606284}.

\bibitem[{\citenamefont{Husek and Leupold}(2015)}]{Husek:2015wta}
\bibinfo{author}{\bibfnamefont{T.}~\bibnamefont{Husek}} \bibnamefont{and}
  \bibinfo{author}{\bibfnamefont{S.}~\bibnamefont{Leupold}},
  \bibinfo{journal}{Eur. Phys. J. C} \textbf{\bibinfo{volume}{75}},
  \bibinfo{pages}{586} (\bibinfo{year}{2015}), \eprint{1507.00478}.

\bibitem[{\citenamefont{Bergstr{\"o}m et~al.}(1983)\citenamefont{Bergstr{\"o}m,
  Mass\'o, Ametller, and Bramon}}]{Bergstrom:1983ay}
\bibinfo{author}{\bibfnamefont{L.}~\bibnamefont{Bergstr{\"o}m}},
  \bibinfo{author}{\bibfnamefont{E.}~\bibnamefont{Mass\'o}},
  \bibinfo{author}{\bibfnamefont{L.}~\bibnamefont{Ametller}}, \bibnamefont{and}
  \bibinfo{author}{\bibfnamefont{A.}~\bibnamefont{Bramon}},
  \bibinfo{journal}{Phys. Lett. B} \textbf{\bibinfo{volume}{126}},
  \bibinfo{pages}{117} (\bibinfo{year}{1983}).

\bibitem[{\citenamefont{Ametller et~al.}(1983)\citenamefont{Ametller,
  Bergstr{\"o}m, Bramon, and Mass{\'o}}}]{Ametller:1983ec}
\bibinfo{author}{\bibfnamefont{L.}~\bibnamefont{Ametller}},
  \bibinfo{author}{\bibfnamefont{L.}~\bibnamefont{Bergstr{\"o}m}},
  \bibinfo{author}{\bibfnamefont{A.}~\bibnamefont{Bramon}}, \bibnamefont{and}
  \bibinfo{author}{\bibfnamefont{E.}~\bibnamefont{Mass{\'o}}},
  \bibinfo{journal}{Nucl. Phys. B} \textbf{\bibinfo{volume}{228}},
  \bibinfo{pages}{301} (\bibinfo{year}{1983}).

\bibitem[{\citenamefont{Dorokhov and Ivanov}(2007)}]{Dorokhov:2007bd}
\bibinfo{author}{\bibfnamefont{A.~E.} \bibnamefont{Dorokhov}} \bibnamefont{and}
  \bibinfo{author}{\bibfnamefont{M.~A.} \bibnamefont{Ivanov}},
  \bibinfo{journal}{Phys. Rev. D} \textbf{\bibinfo{volume}{75}},
  \bibinfo{pages}{114007} (\bibinfo{year}{2007}), \eprint{0704.3498}.

\bibitem[{\citenamefont{Dorokhov and Ivanov}(2008)}]{Dorokhov:2008cd}
\bibinfo{author}{\bibfnamefont{A.~E.} \bibnamefont{Dorokhov}} \bibnamefont{and}
  \bibinfo{author}{\bibfnamefont{M.~A.} \bibnamefont{Ivanov}},
  \bibinfo{journal}{JETP Lett.} \textbf{\bibinfo{volume}{87}},
  \bibinfo{pages}{531} (\bibinfo{year}{2008}), \eprint{0803.4493}.

\bibitem[{\citenamefont{Dorokhov et~al.}(2009)\citenamefont{Dorokhov, Ivanov,
  and Kovalenko}}]{Dorokhov:2009xs}
\bibinfo{author}{\bibfnamefont{A.~E.} \bibnamefont{Dorokhov}},
  \bibinfo{author}{\bibfnamefont{M.~A.} \bibnamefont{Ivanov}},
  \bibnamefont{and} \bibinfo{author}{\bibfnamefont{S.~G.}
  \bibnamefont{Kovalenko}}, \bibinfo{journal}{Phys. Lett. B}
  \textbf{\bibinfo{volume}{677}}, \bibinfo{pages}{145} (\bibinfo{year}{2009}),
  \eprint{0903.4249}.

\bibitem[{\citenamefont{Weil et~al.}(2017)\citenamefont{Weil, Eichmann,
  Fischer, and Williams}}]{Weil:2017knt}
\bibinfo{author}{\bibfnamefont{E.}~\bibnamefont{Weil}},
  \bibinfo{author}{\bibfnamefont{G.}~\bibnamefont{Eichmann}},
  \bibinfo{author}{\bibfnamefont{C.~S.} \bibnamefont{Fischer}},
  \bibnamefont{and} \bibinfo{author}{\bibfnamefont{R.}~\bibnamefont{Williams}},
  \bibinfo{journal}{Phys. Rev. D} \textbf{\bibinfo{volume}{96}},
  \bibinfo{pages}{014021} (\bibinfo{year}{2017}), \eprint{1704.06046}.

\bibitem[{\citenamefont{Eichmann et~al.}(2017)\citenamefont{Eichmann, Fischer,
  Weil, and Williams}}]{Eichmann:2017wil}
\bibinfo{author}{\bibfnamefont{G.}~\bibnamefont{Eichmann}},
  \bibinfo{author}{\bibfnamefont{C.~S.} \bibnamefont{Fischer}},
  \bibinfo{author}{\bibfnamefont{E.}~\bibnamefont{Weil}}, \bibnamefont{and}
  \bibinfo{author}{\bibfnamefont{R.}~\bibnamefont{Williams}},
  \bibinfo{journal}{Phys. Lett. B} \textbf{\bibinfo{volume}{774}},
  \bibinfo{pages}{425} (\bibinfo{year}{2017}), \eprint{1704.05774}.

\bibitem[{\citenamefont{Gronberg et~al.}(1998)}]{Gronberg:1997fj}
\bibinfo{author}{\bibfnamefont{J.}~\bibnamefont{Gronberg}} \bibnamefont{et~al.}
  (\bibinfo{collaboration}{CLEO}), \bibinfo{journal}{Phys. Rev. D}
  \textbf{\bibinfo{volume}{57}}, \bibinfo{pages}{33} (\bibinfo{year}{1998}),
  \eprint{hep-ex/9707031}.

\bibitem[{\citenamefont{Aubert et~al.}(2009)}]{Aubert:2009mc}
\bibinfo{author}{\bibfnamefont{B.}~\bibnamefont{Aubert}} \bibnamefont{et~al.}
  (\bibinfo{collaboration}{BaBar}), \bibinfo{journal}{Phys. Rev. D}
  \textbf{\bibinfo{volume}{80}}, \bibinfo{pages}{052002}
  (\bibinfo{year}{2009}), \eprint{0905.4778}.

\bibitem[{\citenamefont{Uehara et~al.}(2012)}]{Uehara:2012ag}
\bibinfo{author}{\bibfnamefont{S.}~\bibnamefont{Uehara}} \bibnamefont{et~al.}
  (\bibinfo{collaboration}{Belle}), \bibinfo{journal}{Phys. Rev. D}
  \textbf{\bibinfo{volume}{86}}, \bibinfo{pages}{092007}
  (\bibinfo{year}{2012}), \eprint{1205.3249}.

\bibitem[{\citenamefont{Christ et~al.}(2020)\citenamefont{Christ, Feng, Jin,
  Tu, and Zhao}}]{Christ:2020dae}
\bibinfo{author}{\bibfnamefont{N.~H.} \bibnamefont{Christ}},
  \bibinfo{author}{\bibfnamefont{X.}~\bibnamefont{Feng}},
  \bibinfo{author}{\bibfnamefont{L.}~\bibnamefont{Jin}},
  \bibinfo{author}{\bibfnamefont{C.}~\bibnamefont{Tu}}, \bibnamefont{and}
  \bibinfo{author}{\bibfnamefont{Y.}~\bibnamefont{Zhao}},
  \bibinfo{journal}{PoS} \textbf{\bibinfo{volume}{LATTICE2019}},
  \bibinfo{pages}{097} (\bibinfo{year}{2020}), \eprint{2001.05642}.

\bibitem[{\citenamefont{Hoferichter
  et~al.}(2014{\natexlab{b}})\citenamefont{Hoferichter, Colangelo, Procura, and
  Stoffer}}]{Hoferichter:2013ama}
\bibinfo{author}{\bibfnamefont{M.}~\bibnamefont{Hoferichter}},
  \bibinfo{author}{\bibfnamefont{G.}~\bibnamefont{Colangelo}},
  \bibinfo{author}{\bibfnamefont{M.}~\bibnamefont{Procura}}, \bibnamefont{and}
  \bibinfo{author}{\bibfnamefont{P.}~\bibnamefont{Stoffer}},
  \bibinfo{journal}{Int. J. Mod. Phys. Conf. Ser.}
  \textbf{\bibinfo{volume}{35}}, \bibinfo{pages}{1460400}
  (\bibinfo{year}{2014}{\natexlab{b}}), \eprint{1309.6877}.

\bibitem[{\citenamefont{Colangelo
  et~al.}(2014{\natexlab{a}})\citenamefont{Colangelo, Hoferichter, Procura, and
  Stoffer}}]{Colangelo:2014dfa}
\bibinfo{author}{\bibfnamefont{G.}~\bibnamefont{Colangelo}},
  \bibinfo{author}{\bibfnamefont{M.}~\bibnamefont{Hoferichter}},
  \bibinfo{author}{\bibfnamefont{M.}~\bibnamefont{Procura}}, \bibnamefont{and}
  \bibinfo{author}{\bibfnamefont{P.}~\bibnamefont{Stoffer}},
  \bibinfo{journal}{JHEP} \textbf{\bibinfo{volume}{09}}, \bibinfo{pages}{091}
  (\bibinfo{year}{2014}{\natexlab{a}}), \eprint{1402.7081}.

\bibitem[{\citenamefont{Colangelo
  et~al.}(2014{\natexlab{b}})\citenamefont{Colangelo, Hoferichter, Kubis,
  Procura, and Stoffer}}]{Colangelo:2014pva}
\bibinfo{author}{\bibfnamefont{G.}~\bibnamefont{Colangelo}},
  \bibinfo{author}{\bibfnamefont{M.}~\bibnamefont{Hoferichter}},
  \bibinfo{author}{\bibfnamefont{B.}~\bibnamefont{Kubis}},
  \bibinfo{author}{\bibfnamefont{M.}~\bibnamefont{Procura}}, \bibnamefont{and}
  \bibinfo{author}{\bibfnamefont{P.}~\bibnamefont{Stoffer}},
  \bibinfo{journal}{Phys. Lett. B} \textbf{\bibinfo{volume}{738}},
  \bibinfo{pages}{6} (\bibinfo{year}{2014}{\natexlab{b}}), \eprint{1408.2517}.

\bibitem[{\citenamefont{Colangelo et~al.}(2015)\citenamefont{Colangelo,
  Hoferichter, Procura, and Stoffer}}]{Colangelo:2015ama}
\bibinfo{author}{\bibfnamefont{G.}~\bibnamefont{Colangelo}},
  \bibinfo{author}{\bibfnamefont{M.}~\bibnamefont{Hoferichter}},
  \bibinfo{author}{\bibfnamefont{M.}~\bibnamefont{Procura}}, \bibnamefont{and}
  \bibinfo{author}{\bibfnamefont{P.}~\bibnamefont{Stoffer}},
  \bibinfo{journal}{JHEP} \textbf{\bibinfo{volume}{09}}, \bibinfo{pages}{074}
  (\bibinfo{year}{2015}), \eprint{1506.01386}.

\bibitem[{\citenamefont{Colangelo
  et~al.}(2017{\natexlab{a}})\citenamefont{Colangelo, Hoferichter, Procura, and
  Stoffer}}]{Colangelo:2017qdm}
\bibinfo{author}{\bibfnamefont{G.}~\bibnamefont{Colangelo}},
  \bibinfo{author}{\bibfnamefont{M.}~\bibnamefont{Hoferichter}},
  \bibinfo{author}{\bibfnamefont{M.}~\bibnamefont{Procura}}, \bibnamefont{and}
  \bibinfo{author}{\bibfnamefont{P.}~\bibnamefont{Stoffer}},
  \bibinfo{journal}{Phys. Rev. Lett.} \textbf{\bibinfo{volume}{118}},
  \bibinfo{pages}{232001} (\bibinfo{year}{2017}{\natexlab{a}}),
  \eprint{1701.06554}.

\bibitem[{\citenamefont{Colangelo
  et~al.}(2017{\natexlab{b}})\citenamefont{Colangelo, Hoferichter, Procura, and
  Stoffer}}]{Colangelo:2017fiz}
\bibinfo{author}{\bibfnamefont{G.}~\bibnamefont{Colangelo}},
  \bibinfo{author}{\bibfnamefont{M.}~\bibnamefont{Hoferichter}},
  \bibinfo{author}{\bibfnamefont{M.}~\bibnamefont{Procura}}, \bibnamefont{and}
  \bibinfo{author}{\bibfnamefont{P.}~\bibnamefont{Stoffer}},
  \bibinfo{journal}{JHEP} \textbf{\bibinfo{volume}{04}}, \bibinfo{pages}{161}
  (\bibinfo{year}{2017}{\natexlab{b}}), \eprint{1702.07347}.

\bibitem[{\citenamefont{Danilkin et~al.}(2021)\citenamefont{Danilkin,
  Hoferichter, and Stoffer}}]{Danilkin:2021icn}
\bibinfo{author}{\bibfnamefont{I.}~\bibnamefont{Danilkin}},
  \bibinfo{author}{\bibfnamefont{M.}~\bibnamefont{Hoferichter}},
  \bibnamefont{and} \bibinfo{author}{\bibfnamefont{P.}~\bibnamefont{Stoffer}},
  \bibinfo{journal}{Phys. Lett. B} \textbf{\bibinfo{volume}{820}},
  \bibinfo{pages}{136502} (\bibinfo{year}{2021}), \eprint{2105.01666}.

\bibitem[{\citenamefont{Hoferichter et~al.}(2019)\citenamefont{Hoferichter,
  Hoid, and Kubis}}]{Hoferichter:2019mqg}
\bibinfo{author}{\bibfnamefont{M.}~\bibnamefont{Hoferichter}},
  \bibinfo{author}{\bibfnamefont{B.-L.} \bibnamefont{Hoid}}, \bibnamefont{and}
  \bibinfo{author}{\bibfnamefont{B.}~\bibnamefont{Kubis}},
  \bibinfo{journal}{JHEP} \textbf{\bibinfo{volume}{08}}, \bibinfo{pages}{137}
  (\bibinfo{year}{2019}), \eprint{1907.01556}.

\bibitem[{\citenamefont{Hoid et~al.}(2020)\citenamefont{Hoid, Hoferichter, and
  Kubis}}]{Hoid:2020xjs}
\bibinfo{author}{\bibfnamefont{B.-L.} \bibnamefont{Hoid}},
  \bibinfo{author}{\bibfnamefont{M.}~\bibnamefont{Hoferichter}},
  \bibnamefont{and} \bibinfo{author}{\bibfnamefont{B.}~\bibnamefont{Kubis}},
  \bibinfo{journal}{Eur. Phys. J. C} \textbf{\bibinfo{volume}{80}},
  \bibinfo{pages}{988} (\bibinfo{year}{2020}), \eprint{2007.12696}.

\bibitem[{\citenamefont{Colangelo
  et~al.}(2020{\natexlab{a}})\citenamefont{Colangelo, Hagelstein, Hoferichter,
  Laub, and Stoffer}}]{Colangelo:2019lpu}
\bibinfo{author}{\bibfnamefont{G.}~\bibnamefont{Colangelo}},
  \bibinfo{author}{\bibfnamefont{F.}~\bibnamefont{Hagelstein}},
  \bibinfo{author}{\bibfnamefont{M.}~\bibnamefont{Hoferichter}},
  \bibinfo{author}{\bibfnamefont{L.}~\bibnamefont{Laub}}, \bibnamefont{and}
  \bibinfo{author}{\bibfnamefont{P.}~\bibnamefont{Stoffer}},
  \bibinfo{journal}{Phys. Rev. D} \textbf{\bibinfo{volume}{101}},
  \bibinfo{pages}{051501} (\bibinfo{year}{2020}{\natexlab{a}}),
  \eprint{1910.11881}.

\bibitem[{\citenamefont{Colangelo
  et~al.}(2020{\natexlab{b}})\citenamefont{Colangelo, Hagelstein, Hoferichter,
  Laub, and Stoffer}}]{Colangelo:2019uex}
\bibinfo{author}{\bibfnamefont{G.}~\bibnamefont{Colangelo}},
  \bibinfo{author}{\bibfnamefont{F.}~\bibnamefont{Hagelstein}},
  \bibinfo{author}{\bibfnamefont{M.}~\bibnamefont{Hoferichter}},
  \bibinfo{author}{\bibfnamefont{L.}~\bibnamefont{Laub}}, \bibnamefont{and}
  \bibinfo{author}{\bibfnamefont{P.}~\bibnamefont{Stoffer}},
  \bibinfo{journal}{JHEP} \textbf{\bibinfo{volume}{03}}, \bibinfo{pages}{101}
  (\bibinfo{year}{2020}{\natexlab{b}}), \eprint{1910.13432}.

\bibitem[{\citenamefont{Hoferichter and Stoffer}(2020)}]{Hoferichter:2020lap}
\bibinfo{author}{\bibfnamefont{M.}~\bibnamefont{Hoferichter}} \bibnamefont{and}
  \bibinfo{author}{\bibfnamefont{P.}~\bibnamefont{Stoffer}},
  \bibinfo{journal}{JHEP} \textbf{\bibinfo{volume}{05}}, \bibinfo{pages}{159}
  (\bibinfo{year}{2020}), \eprint{2004.06127}.

\bibitem[{\citenamefont{G\'erardin et~al.}(2019)\citenamefont{G\'erardin,
  Meyer, and Nyffeler}}]{Gerardin:2019vio}
\bibinfo{author}{\bibfnamefont{A.}~\bibnamefont{G\'erardin}},
  \bibinfo{author}{\bibfnamefont{H.~B.} \bibnamefont{Meyer}}, \bibnamefont{and}
  \bibinfo{author}{\bibfnamefont{A.}~\bibnamefont{Nyffeler}},
  \bibinfo{journal}{Phys. Rev. D} \textbf{\bibinfo{volume}{100}},
  \bibinfo{pages}{034520} (\bibinfo{year}{2019}), \eprint{1903.09471}.

\bibitem[{\citenamefont{Knecht and Nyffeler}(2002)}]{Knecht:2001qf}
\bibinfo{author}{\bibfnamefont{M.}~\bibnamefont{Knecht}} \bibnamefont{and}
  \bibinfo{author}{\bibfnamefont{A.}~\bibnamefont{Nyffeler}},
  \bibinfo{journal}{Phys. Rev. D} \textbf{\bibinfo{volume}{65}},
  \bibinfo{pages}{073034} (\bibinfo{year}{2002}), \eprint{hep-ph/0111058}.

\bibitem[{\citenamefont{Lucha et~al.}(2007)\citenamefont{Lucha, Melikhov, and
  Simula}}]{Lucha:2006vc}
\bibinfo{author}{\bibfnamefont{W.}~\bibnamefont{Lucha}},
  \bibinfo{author}{\bibfnamefont{D.}~\bibnamefont{Melikhov}}, \bibnamefont{and}
  \bibinfo{author}{\bibfnamefont{S.}~\bibnamefont{Simula}},
  \bibinfo{journal}{Phys. Rev. D} \textbf{\bibinfo{volume}{75}},
  \bibinfo{pages}{016001} (\bibinfo{year}{2007}), \bibinfo{note}{[Erratum:
  Phys. Rev. D {\bf 92}, 019901 (2015)]}, \eprint{hep-ph/0610330}.

\bibitem[{\citenamefont{Colangelo et~al.}(2019)\citenamefont{Colangelo,
  Hoferichter, and Stoffer}}]{Colangelo:2018mtw}
\bibinfo{author}{\bibfnamefont{G.}~\bibnamefont{Colangelo}},
  \bibinfo{author}{\bibfnamefont{M.}~\bibnamefont{Hoferichter}},
  \bibnamefont{and} \bibinfo{author}{\bibfnamefont{P.}~\bibnamefont{Stoffer}},
  \bibinfo{journal}{JHEP} \textbf{\bibinfo{volume}{02}}, \bibinfo{pages}{006}
  (\bibinfo{year}{2019}), \eprint{1810.00007}.

\bibitem[{\citenamefont{Colangelo et~al.}(2021)\citenamefont{Colangelo,
  Hoferichter, and Stoffer}}]{Colangelo:2020lcg}
\bibinfo{author}{\bibfnamefont{G.}~\bibnamefont{Colangelo}},
  \bibinfo{author}{\bibfnamefont{M.}~\bibnamefont{Hoferichter}},
  \bibnamefont{and} \bibinfo{author}{\bibfnamefont{P.}~\bibnamefont{Stoffer}},
  \bibinfo{journal}{Phys. Lett. B} \textbf{\bibinfo{volume}{814}},
  \bibinfo{pages}{136073} (\bibinfo{year}{2021}), \eprint{2010.07943}.

\bibitem[{\citenamefont{Khuri and Treiman}(1960)}]{Khuri:1960zz}
\bibinfo{author}{\bibfnamefont{N.~N.} \bibnamefont{Khuri}} \bibnamefont{and}
  \bibinfo{author}{\bibfnamefont{S.~B.} \bibnamefont{Treiman}},
  \bibinfo{journal}{Phys. Rev.} \textbf{\bibinfo{volume}{119}},
  \bibinfo{pages}{1115} (\bibinfo{year}{1960}).

\bibitem[{\citenamefont{Lepage and Brodsky}(1979)}]{Lepage:1979zb}
\bibinfo{author}{\bibfnamefont{G.~P.} \bibnamefont{Lepage}} \bibnamefont{and}
  \bibinfo{author}{\bibfnamefont{S.~J.} \bibnamefont{Brodsky}},
  \bibinfo{journal}{Phys. Lett. B} \textbf{\bibinfo{volume}{87}},
  \bibinfo{pages}{359} (\bibinfo{year}{1979}).

\bibitem[{\citenamefont{Lepage and Brodsky}(1980)}]{Lepage:1980fj}
\bibinfo{author}{\bibfnamefont{G.~P.} \bibnamefont{Lepage}} \bibnamefont{and}
  \bibinfo{author}{\bibfnamefont{S.~J.} \bibnamefont{Brodsky}},
  \bibinfo{journal}{Phys. Rev. D} \textbf{\bibinfo{volume}{22}},
  \bibinfo{pages}{2157} (\bibinfo{year}{1980}).

\bibitem[{\citenamefont{Khodjamirian}(1999)}]{Khodjamirian:1997tk}
\bibinfo{author}{\bibfnamefont{A.}~\bibnamefont{Khodjamirian}},
  \bibinfo{journal}{Eur. Phys. J. C} \textbf{\bibinfo{volume}{6}},
  \bibinfo{pages}{477} (\bibinfo{year}{1999}), \eprint{hep-ph/9712451}.

\bibitem[{\citenamefont{Zanke et~al.}(2021)\citenamefont{Zanke, Hoferichter,
  and Kubis}}]{Zanke:2021wiq}
\bibinfo{author}{\bibfnamefont{M.}~\bibnamefont{Zanke}},
  \bibinfo{author}{\bibfnamefont{M.}~\bibnamefont{Hoferichter}},
  \bibnamefont{and} \bibinfo{author}{\bibfnamefont{B.}~\bibnamefont{Kubis}},
  \bibinfo{journal}{JHEP} \textbf{\bibinfo{volume}{07}}, \bibinfo{pages}{106}
  (\bibinfo{year}{2021}), \eprint{2103.09829}.

\bibitem[{\citenamefont{Agaev et~al.}(2011)\citenamefont{Agaev, Braun, Offen,
  and Porkert}}]{Agaev:2010aq}
\bibinfo{author}{\bibfnamefont{S.~S.} \bibnamefont{Agaev}},
  \bibinfo{author}{\bibfnamefont{V.~M.} \bibnamefont{Braun}},
  \bibinfo{author}{\bibfnamefont{N.}~\bibnamefont{Offen}}, \bibnamefont{and}
  \bibinfo{author}{\bibfnamefont{F.~A.} \bibnamefont{Porkert}},
  \bibinfo{journal}{Phys. Rev. D} \textbf{\bibinfo{volume}{83}},
  \bibinfo{pages}{054020} (\bibinfo{year}{2011}), \eprint{1012.4671}.

\bibitem[{\citenamefont{Mikhailov et~al.}(2016)\citenamefont{Mikhailov,
  Pimikov, and Stefanis}}]{Mikhailov:2016klg}
\bibinfo{author}{\bibfnamefont{S.~V.} \bibnamefont{Mikhailov}},
  \bibinfo{author}{\bibfnamefont{A.~V.} \bibnamefont{Pimikov}},
  \bibnamefont{and} \bibinfo{author}{\bibfnamefont{N.~G.}
  \bibnamefont{Stefanis}}, \bibinfo{journal}{Phys. Rev. D}
  \textbf{\bibinfo{volume}{93}}, \bibinfo{pages}{114018}
  (\bibinfo{year}{2016}), \eprint{1604.06391}.

\bibitem[{Sup()}]{Supparxiv}
\bibinfo{note}{See appendix for more details on the evaluation of the loop
  integrals.}

\bibitem[{\citenamefont{L{\"u}dtke}(2016)}]{Luedtke:2016}
\bibinfo{author}{\bibfnamefont{J.}~\bibnamefont{L{\"u}dtke}},
  \bibinfo{type}{Bachelor's thesis}, \bibinfo{school}{University of Bonn}
  (\bibinfo{year}{2016}).

\bibitem[{\citenamefont{Hoid}(2020)}]{Hoid:2020qij}
\bibinfo{author}{\bibfnamefont{B.-L.} \bibnamefont{Hoid}}, Ph.D. thesis,
  \bibinfo{school}{University of Bonn} (\bibinfo{year}{2020}),
  \urlprefix\url{https://hdl.handle.net/20.500.11811/8906}.

\bibitem[{\citenamefont{Rosner}(1967)}]{Rosner:1967zz}
\bibinfo{author}{\bibfnamefont{J.~L.} \bibnamefont{Rosner}},
  \bibinfo{journal}{Annals Phys.} \textbf{\bibinfo{volume}{44}},
  \bibinfo{pages}{11} (\bibinfo{year}{1967}).

\bibitem[{\citenamefont{Levine and Roskies}(1974)}]{Levine:1974xh}
\bibinfo{author}{\bibfnamefont{M.~J.} \bibnamefont{Levine}} \bibnamefont{and}
  \bibinfo{author}{\bibfnamefont{R.}~\bibnamefont{Roskies}},
  \bibinfo{journal}{Phys. Rev. D} \textbf{\bibinfo{volume}{9}},
  \bibinfo{pages}{421} (\bibinfo{year}{1974}).

\bibitem[{\citenamefont{Levine et~al.}(1979)\citenamefont{Levine, Remiddi, and
  Roskies}}]{Levine:1979uz}
\bibinfo{author}{\bibfnamefont{M.~J.} \bibnamefont{Levine}},
  \bibinfo{author}{\bibfnamefont{E.}~\bibnamefont{Remiddi}}, \bibnamefont{and}
  \bibinfo{author}{\bibfnamefont{R.}~\bibnamefont{Roskies}},
  \bibinfo{journal}{Phys. Rev. D} \textbf{\bibinfo{volume}{20}},
  \bibinfo{pages}{2068} (\bibinfo{year}{1979}).

\bibitem[{\citenamefont{'t~Hooft and Veltman}(1979)}]{tHooft:1978jhc}
\bibinfo{author}{\bibfnamefont{G.}~\bibnamefont{'t~Hooft}} \bibnamefont{and}
  \bibinfo{author}{\bibfnamefont{M.~J.~G.} \bibnamefont{Veltman}},
  \bibinfo{journal}{Nucl. Phys. B} \textbf{\bibinfo{volume}{153}},
  \bibinfo{pages}{365} (\bibinfo{year}{1979}).

\bibitem[{\citenamefont{Hahn and P{\'e}rez-Victoria}(1999)}]{Hahn:1998yk}
\bibinfo{author}{\bibfnamefont{T.}~\bibnamefont{Hahn}} \bibnamefont{and}
  \bibinfo{author}{\bibfnamefont{M.}~\bibnamefont{P{\'e}rez-Victoria}},
  \bibinfo{journal}{Comput. Phys. Commun.} \textbf{\bibinfo{volume}{118}},
  \bibinfo{pages}{153} (\bibinfo{year}{1999}), \eprint{hep-ph/9807565}.

\bibitem[{\citenamefont{S{\'a}nchez-Puertas}(2019)}]{Sanchez-Puertas:2018tnp}
\bibinfo{author}{\bibfnamefont{P.}~\bibnamefont{S{\'a}nchez-Puertas}},
  \bibinfo{journal}{JHEP} \textbf{\bibinfo{volume}{01}}, \bibinfo{pages}{031}
  (\bibinfo{year}{2019}), \eprint{1810.13228}.

\bibitem[{\citenamefont{Aoki et~al.}(2020)}]{Aoki:2019cca}
\bibinfo{author}{\bibfnamefont{S.}~\bibnamefont{Aoki}} \bibnamefont{et~al.}
  (\bibinfo{collaboration}{Flavour Lattice Averaging Group}),
  \bibinfo{journal}{Eur. Phys. J. C} \textbf{\bibinfo{volume}{80}},
  \bibinfo{pages}{113} (\bibinfo{year}{2020}), \eprint{1902.08191}.

\bibitem[{\citenamefont{McNeile et~al.}(2010)\citenamefont{McNeile, Davies,
  Follana, Hornbostel, and Lepage}}]{McNeile:2010ji}
\bibinfo{author}{\bibfnamefont{C.}~\bibnamefont{McNeile}},
  \bibinfo{author}{\bibfnamefont{C.~T.~H.} \bibnamefont{Davies}},
  \bibinfo{author}{\bibfnamefont{E.}~\bibnamefont{Follana}},
  \bibinfo{author}{\bibfnamefont{K.}~\bibnamefont{Hornbostel}},
  \bibnamefont{and} \bibinfo{author}{\bibfnamefont{G.~P.}
  \bibnamefont{Lepage}}, \bibinfo{journal}{Phys. Rev. D}
  \textbf{\bibinfo{volume}{82}}, \bibinfo{pages}{034512}
  (\bibinfo{year}{2010}), \eprint{1004.4285}.

\bibitem[{\citenamefont{D{\"u}rr et~al.}(2011)}]{Durr:2010vn}
\bibinfo{author}{\bibfnamefont{S.}~\bibnamefont{D{\"u}rr}}
  \bibnamefont{et~al.}, \bibinfo{journal}{Phys. Lett. B}
  \textbf{\bibinfo{volume}{701}}, \bibinfo{pages}{265} (\bibinfo{year}{2011}),
  \eprint{1011.2403}.

\bibitem[{\citenamefont{Carrasco et~al.}(2014)}]{Carrasco:2014cwa}
\bibinfo{author}{\bibfnamefont{N.}~\bibnamefont{Carrasco}} \bibnamefont{et~al.}
  (\bibinfo{collaboration}{European Twisted Mass}), \bibinfo{journal}{Nucl.
  Phys. B} \textbf{\bibinfo{volume}{887}}, \bibinfo{pages}{19}
  (\bibinfo{year}{2014}), \eprint{1403.4504}.

\bibitem[{\citenamefont{Blum et~al.}(2016)}]{Blum:2014tka}
\bibinfo{author}{\bibfnamefont{T.}~\bibnamefont{Blum}} \bibnamefont{et~al.}
  (\bibinfo{collaboration}{RBC, UKQCD}), \bibinfo{journal}{Phys. Rev. D}
  \textbf{\bibinfo{volume}{93}}, \bibinfo{pages}{074505}
  (\bibinfo{year}{2016}), \eprint{1411.7017}.

\bibitem[{\citenamefont{Bazavov et~al.}(2018)}]{Bazavov:2018omf}
\bibinfo{author}{\bibfnamefont{A.}~\bibnamefont{Bazavov}} \bibnamefont{et~al.}
  (\bibinfo{collaboration}{Fermilab Lattice, MILC, TUMQCD}),
  \bibinfo{journal}{Phys. Rev. D} \textbf{\bibinfo{volume}{98}},
  \bibinfo{pages}{054517} (\bibinfo{year}{2018}), \eprint{1802.04248}.

\bibitem[{\citenamefont{Grzadkowski et~al.}(2010)\citenamefont{Grzadkowski,
  Iskrzy{\'n}ski, Misiak, and Rosiek}}]{Grzadkowski:2010es}
\bibinfo{author}{\bibfnamefont{B.}~\bibnamefont{Grzadkowski}},
  \bibinfo{author}{\bibfnamefont{M.}~\bibnamefont{Iskrzy{\'n}ski}},
  \bibinfo{author}{\bibfnamefont{M.}~\bibnamefont{Misiak}}, \bibnamefont{and}
  \bibinfo{author}{\bibfnamefont{J.}~\bibnamefont{Rosiek}},
  \bibinfo{journal}{JHEP} \textbf{\bibinfo{volume}{10}}, \bibinfo{pages}{085}
  (\bibinfo{year}{2010}), \eprint{1008.4884}.

\bibitem[{\citenamefont{Buchm{\"u}ller and Wyler}(1986)}]{Buchmuller:1985jz}
\bibinfo{author}{\bibfnamefont{W.}~\bibnamefont{Buchm{\"u}ller}}
  \bibnamefont{and} \bibinfo{author}{\bibfnamefont{D.}~\bibnamefont{Wyler}},
  \bibinfo{journal}{Nucl. Phys. B} \textbf{\bibinfo{volume}{268}},
  \bibinfo{pages}{621} (\bibinfo{year}{1986}).

\bibitem[{\citenamefont{de~Blas et~al.}(2013)\citenamefont{de~Blas, Chala, and
  Santiago}}]{deBlas:2013qqa}
\bibinfo{author}{\bibfnamefont{J.}~\bibnamefont{de~Blas}},
  \bibinfo{author}{\bibfnamefont{M.}~\bibnamefont{Chala}}, \bibnamefont{and}
  \bibinfo{author}{\bibfnamefont{J.}~\bibnamefont{Santiago}},
  \bibinfo{journal}{Phys. Rev. D} \textbf{\bibinfo{volume}{88}},
  \bibinfo{pages}{095011} (\bibinfo{year}{2013}), \eprint{1307.5068}.

\bibitem[{\citenamefont{Falkowski et~al.}(2017)\citenamefont{Falkowski,
  Gonz\'alez-Alonso, and Mimouni}}]{Falkowski:2017pss}
\bibinfo{author}{\bibfnamefont{A.}~\bibnamefont{Falkowski}},
  \bibinfo{author}{\bibfnamefont{M.}~\bibnamefont{Gonz\'alez-Alonso}},
  \bibnamefont{and} \bibinfo{author}{\bibfnamefont{K.}~\bibnamefont{Mimouni}},
  \bibinfo{journal}{JHEP} \textbf{\bibinfo{volume}{08}}, \bibinfo{pages}{123}
  (\bibinfo{year}{2017}), \eprint{1706.03783}.

\bibitem[{\citenamefont{Crivellin et~al.}(2021)\citenamefont{Crivellin,
  Hoferichter, Kirk, Manzari, and Schnell}}]{Crivellin:2021bkd}
\bibinfo{author}{\bibfnamefont{A.}~\bibnamefont{Crivellin}},
  \bibinfo{author}{\bibfnamefont{M.}~\bibnamefont{Hoferichter}},
  \bibinfo{author}{\bibfnamefont{M.}~\bibnamefont{Kirk}},
  \bibinfo{author}{\bibfnamefont{C.~A.} \bibnamefont{Manzari}},
  \bibnamefont{and} \bibinfo{author}{\bibfnamefont{L.}~\bibnamefont{Schnell}},
  \bibinfo{journal}{JHEP} \textbf{\bibinfo{volume}{10}}, \bibinfo{pages}{221}
  (\bibinfo{year}{2021}), \eprint{2107.13569}.

\bibitem[{\citenamefont{Hanneke et~al.}(2008)\citenamefont{Hanneke, Fogwell,
  and Gabrielse}}]{Hanneke:2008tm}
\bibinfo{author}{\bibfnamefont{D.}~\bibnamefont{Hanneke}},
  \bibinfo{author}{\bibfnamefont{S.}~\bibnamefont{Fogwell}}, \bibnamefont{and}
  \bibinfo{author}{\bibfnamefont{G.}~\bibnamefont{Gabrielse}},
  \bibinfo{journal}{Phys. Rev. Lett.} \textbf{\bibinfo{volume}{100}},
  \bibinfo{pages}{120801} (\bibinfo{year}{2008}), \eprint{0801.1134}.

\bibitem[{\citenamefont{Aoyama et~al.}(2019)\citenamefont{Aoyama, Kinoshita,
  and Nio}}]{Aoyama:2019ryr}
\bibinfo{author}{\bibfnamefont{T.}~\bibnamefont{Aoyama}},
  \bibinfo{author}{\bibfnamefont{T.}~\bibnamefont{Kinoshita}},
  \bibnamefont{and} \bibinfo{author}{\bibfnamefont{M.}~\bibnamefont{Nio}},
  \bibinfo{journal}{Atoms} \textbf{\bibinfo{volume}{7}}, \bibinfo{pages}{28}
  (\bibinfo{year}{2019}).

\bibitem[{\citenamefont{Aoyama et~al.}(2020)}]{Aoyama:2020ynm}
\bibinfo{author}{\bibfnamefont{T.}~\bibnamefont{Aoyama}} \bibnamefont{et~al.},
  \bibinfo{journal}{Phys. Rept.} \textbf{\bibinfo{volume}{887}},
  \bibinfo{pages}{1} (\bibinfo{year}{2020}), \eprint{2006.04822}.

\bibitem[{\citenamefont{Parker et~al.}(2018)\citenamefont{Parker, Yu, Zhong,
  Estey, and M\"uller}}]{Parker:2018vye}
\bibinfo{author}{\bibfnamefont{R.~H.} \bibnamefont{Parker}},
  \bibinfo{author}{\bibfnamefont{C.}~\bibnamefont{Yu}},
  \bibinfo{author}{\bibfnamefont{W.}~\bibnamefont{Zhong}},
  \bibinfo{author}{\bibfnamefont{B.}~\bibnamefont{Estey}}, \bibnamefont{and}
  \bibinfo{author}{\bibfnamefont{H.}~\bibnamefont{M\"uller}},
  \bibinfo{journal}{Science} \textbf{\bibinfo{volume}{360}},
  \bibinfo{pages}{191} (\bibinfo{year}{2018}), \eprint{1812.04130}.

\bibitem[{\citenamefont{Morel et~al.}(2020)\citenamefont{Morel, Yao, Clad\'e,
  and Guellati-Kh\'elifa}}]{Morel:2020dww}
\bibinfo{author}{\bibfnamefont{L.}~\bibnamefont{Morel}},
  \bibinfo{author}{\bibfnamefont{Z.}~\bibnamefont{Yao}},
  \bibinfo{author}{\bibfnamefont{P.}~\bibnamefont{Clad\'e}}, \bibnamefont{and}
  \bibinfo{author}{\bibfnamefont{S.}~\bibnamefont{Guellati-Kh\'elifa}},
  \bibinfo{journal}{Nature} \textbf{\bibinfo{volume}{588}}, \bibinfo{pages}{61}
  (\bibinfo{year}{2020}).

\bibitem[{\citenamefont{Leveille}(1978)}]{Leveille:1977rc}
\bibinfo{author}{\bibfnamefont{J.~P.} \bibnamefont{Leveille}},
  \bibinfo{journal}{Nucl. Phys. B} \textbf{\bibinfo{volume}{137}},
  \bibinfo{pages}{63} (\bibinfo{year}{1978}).

\bibitem[{\citenamefont{Liu et~al.}(2019)\citenamefont{Liu, Wagner, and
  Wang}}]{Liu:2018xkx}
\bibinfo{author}{\bibfnamefont{J.}~\bibnamefont{Liu}},
  \bibinfo{author}{\bibfnamefont{C.~E.~M.} \bibnamefont{Wagner}},
  \bibnamefont{and} \bibinfo{author}{\bibfnamefont{X.-P.} \bibnamefont{Wang}},
  \bibinfo{journal}{JHEP} \textbf{\bibinfo{volume}{03}}, \bibinfo{pages}{008}
  (\bibinfo{year}{2019}), \eprint{1810.11028}.

\bibitem[{\citenamefont{Lees et~al.}(2017)}]{Lees:2017lec}
\bibinfo{author}{\bibfnamefont{J.~P.} \bibnamefont{Lees}} \bibnamefont{et~al.}
  (\bibinfo{collaboration}{BaBar}), \bibinfo{journal}{Phys. Rev. Lett.}
  \textbf{\bibinfo{volume}{119}}, \bibinfo{pages}{131804}
  (\bibinfo{year}{2017}), \eprint{1702.03327}.

\bibitem[{\citenamefont{NA62}(2020)}]{NA62:2713499}
\bibinfo{author}{\bibnamefont{NA62}}, \emph{\bibinfo{title}{{2020 NA62 Status
  Report to the CERN SPSC}}},
  \bibinfo{howpublished}{\url{https://cds.cern.ch/record/2713499}}
  (\bibinfo{year}{2020}).

\end{thebibliography}

\end{document}